%% file: mpdiam_sisc.tex
\documentclass[final]{siamltex}
\usepackage{mathrsfs}
\usepackage{amsmath,amssymb}
\usepackage{color} 
\usepackage{graphicx}
\usepackage[english]{babel}
\usepackage{algorithm,algorithmic}
\usepackage[notref, notcite]{showkeys}
\usepackage{makros}
\usepackage{listings}
\DeclareGraphicsExtensions{.pdf,.jpeg,.png}
\usepackage{subfigure}

\usepackage{algorithm}
\usepackage{algorithmic}

\lstdefinestyle{mystyle}{
  language=C,
  stepnumber=1,
  numbersep=10pt,
  tabsize=2,
  showspaces=false,
  showstringspaces=false,
  breaklines=true
}

   \usepackage{ifthen,xkeyval,xfor,amsgen}
   \usepackage[acronym,toc]{glossaries}
   \newglossary[slg]{symbols}{syi}{sbl}{List of Symbols}
   \makeglossaries
   \include{Lists}

\title{
Accelerated dimension-independent adaptive Metropolis} 

\author{Yuxin Chen\thanks{Applied Mathematics and Computational Sciences and Extreme Computing Research Center, KAUST,
  Thuwal, Saudi Arabia ({\tt yuxin.chen@kaust.edu.sa})}
\and
David Keyes\thanks{Applied Mathematics and Computational Sciences and Extreme Computing Research Center, KAUST,
  Thuwal, Saudi Arabia ({\tt david.keyes@kaust.edu.sa})}
\and
Kody J.H. Law\thanks{SRI-UQ Center, CEMSE, KAUST, Thuwal, KSA ({\tt kody.law@kaust.edu.sa})}
\and 
Hatem Ltaief\thanks{Applied Mathematics and Computational Sciences and Extreme Computing Research Center, KAUST,
  Thuwal, Saudi Arabia ({\tt hatem.ltaief@kaust.edu.sa})}
}

\begin{document}

\maketitle

\begin{abstract}
This work considers black-box Bayesian inference over high-dimensional parameter spaces. The well-known 
adaptive Metropolis (AM) algorithm  \cite{haario2001adaptive} is extended herein to scale asymptotically uniformly with respect to the underlying parameter dimension for Gaussian targets, by respecting the variance of the target. The resulting algorithm, referred to as the dimension-independent adaptive Metropolis (DIAM) algorithm, also shows improved performance with respect to adaptive Metropolis on non-Gaussian targets. This algorithm is further improved, and the possibility of probing high-dimensional targets is enabled, via GPU-accelerated numerical libraries and periodically synchronized concurrent chains (justified a posteriori). Asymptotically in dimension, this 
GPU implementation exhibits a factor of four improvement versus 
a competitive CPU-based Intel MKL parallel version alone.
Strong scaling to concurrent chains 
is exhibited, through a combination of longer time per sample batch (weak scaling) and yet fewer necessary samples to convergence. 
The algorithm performance is illustrated 
on several Gaussian and non-Gaussian target examples, in which the 
dimension may be in excess of one thousand.

\end{abstract}


\pagestyle{myheadings}
\thispagestyle{plain}
\markboth{Accelerated DIAM}{Accelerated DIAM}

\input{Introduction}

\input{Chapter_2}

\input{Chapter_3}

\input{Chapter_4}
\input{Chapter_5}

\input{Conclusion}

{\bf Acknowledgements } Research reported in this publication was supported by the King Abdullah University of Science and Technology (KAUST).  YC, DK, and HL are members of the Extreme Computing Research Center at KAUST.  KJHL is a member of the SRI Center for Uncertainty Quantification at KAUST.

\bibliography{mybib}
\bibliographystyle{plain}

\end{document}

%% file: Lists.tex
\newglossaryentry{symb:P}{
name=$P$, type=symbols,
description=Number of chains run concurrently,
sort=symbolp
}

\newglossaryentry{symb:M}{
name=$M$, type=symbols,
description=Number of proposal updates done within a batch,
sort=symbolm
}

\newglossaryentry{symb:nlag}{
name=$n_{\rm lag}$, type=symbols,
description={Number of samples generated to update proposal, decides frequency we update the proposal},
sort=symbolnlag
}

\newglossaryentry{symb:sl}{
name=$S^p_{k,m}$, type=symbols,
description={Local second moment to compute the accumulated second moment within current batch, with subscriptions k denotes number of batches we have done and m denotes number of proposal update we have done within current batch },
sort=symbolsl
}

\newglossaryentry{symb:ml}{
name=$m^p_{k,m}$, type=symbols,
description={Local mean compute the accumulated mean within current batch with subscriptions k denotes number of batches we have done and m denotes number of proposal update we have done within current batch},
sort=symbolml
}

\newglossaryentry{symb:sg}{
name=$S^{p,{\rm glob}}_{k,m}$, type=symbols,
description={Global second moment to compute the accumulated second moment with subscriptions p denotes number of chains we run concurrently, k denotes number of batches we have done and m denotes number of proposal update we have done within current batch},
sort=symbolsg
}

\newglossaryentry{symb:mg}{
name=$m^{p,{\rm glob}}_{k,m}$, type=symbols,
description={Global mean to compute the accumulated mean with subscriptions p denotes number of chains we run concurrently, k denotes number of batches we have done and m denotes number of proposal update we have done within current batch},
sort=symbolmg
}

\newglossaryentry{symb:cg}{
name=$C^{p,{\rm glob}}_{k,m}$, type=symbols,
description={This is the component $C_n$ we use to construct MCMC proposal: $x_{n+1} = x_{\rm ref} + \sqrt{1-\beta^2} (x_n -  x_{\rm ref}) + \beta C_n W_n$. it calculated from accumulated sample covariance with subscriptions p denotes number of chains we run concurrently, k denotes number of batches we have done and m denotes number of proposal update we have done within the current batch  },
sort=symbolcg
}

\newglossaryentry{symb:sc}{
name=$S^{{\rm glob}}_k$, type=symbols,
description={Global second moment base to compute the accumulated second moment to the latest finished batch with subscription k denotes number of batches we have done completely and not include the batch we are working on and p denotes number of chains we run concurrently },
sort=symbolsc
}

\newglossaryentry{symb:mc}{
name=$m^{{\rm glob}}_k$, type=symbols,
description={Global mean base to compute the accumulated second moment to the latest finished batch with subscription k denotes number of batches we have done completely and not include the batch we are working on and p denotes the number of chains we run concurrently  },
sort=symbolmc
}

\newacronym{am}{AM}{adaptive Metropolis algorithm}
\newacronym{ode}{ODE}{ordinary differential equation}
\newacronym{pde}{PDE}{partial differential equation}
\newacronym{diam}{DIAM}{dimension-independent adaptive Metropolis algorithm}
\newacronym{mpdiam}{MPDIAM}{massively parallel dimension-independent adaptive Metropolis algorithm}
\newacronym{mcmc}{MCMC}{Markov chain Monte Carlo}
\newacronym{dram}{DRAM}{}
\newacronym{aswam}{ASWAM}{}
\newacronym{scam}{SCAM}{}
\newacronym{ram}{RAM}{}
\newacronym{mala}{MALA}{Metropolis-Adjusted Langevin algorithm}
\newacronym{hmc}{HMC}{Hamiltonian Monte Carlo}
\newacronym{pcn}{pCN}{pre-conditioned Crank-Nicolson}
\newacronym{rmse}{RMSE}{root mean squared error}
\newacronym{psrf}{PSRF}{potential scale reduction factor}
\newacronym{blas2}{BLAS2}{Basic Linear Algebra Subprograms level-2}
\newacronym{blas3}{BLAS3}{Basic Linear Algebra Subprograms level-3}
\newacronym{blas}{BLAS}{Basic Linear Algebra Subprograms}
\newacronym{mpi}{MPI}{message passing interface}
\newacronym{mh}{MH}{Metropolis-Hastings algorithm}
\newacronym{rw}{RW}{random walk Metropolis algorithm}
\newacronym{sde}{SDE}{stochastic differential equation}
\newacronym{dili}{DILI}{dimension-independent likelihood-informed}
\newacronym{kl}{KL}{Kullback-Liebler}
\newacronym{map}{MAP}{maximum a posteriori estimator}
\newacronym{gm}{GM}{GPU and MKL combined code}
\newacronym{mkl}{MKL}{MKL code}
\newacronym{sc}{SEQUENTIAL C}{Sequential code writen in C}

\newacronym{toc}{ToC}{Table of Contents}
\newacronym{los}{LoS}{List of Symbols}
\newacronym{loa}{AM}{Adaptive Metropolis algorithm}
\newacronym{phd}{PhD}{Doctoral}
\newacronym{MS}{MS}{Masters}
\newacronym{M$}{MS}{Microsoft}
\newacronym{CD}{CD}{Compact Disc}
\newacronym{kaust}{AM}{Adaptive Metropolis algorithm}

\newacronym{AD}{AD}{Active Directory\protect\glsadd{glos:AD}}

\newglossaryentry{glos:AD}{
name=Active Directory,
description={Active Directory is the directory service for Windows based networks, that allows central organization and administration of any network resource. It allows a single-sign-on concept independent from network topologies or network protocols. As a prerequisite you need a Windows Server acting as Domain Controller. This computer stores all necessary data, e.g.~usernames and corresponding passwords}
}

\newglossaryentry{glos:RespF}{name=response file, description={A file 
that allows unattended software installation}}

%% file: Introduction.tex

\section{Introduction}
\label{sec:intro}

Recent years have seen 
increasing 
activity in the areas of uncertainty quantification
and big data, largely enabled by the progress of computational science, which itself is enabled by ever
more powerful computers and the symbiosis of this architectural brute force with innovative algorithmic advances.
In particular, the solution of a forward problem, given by an \gls{ode} or \gls{pde}, 
may be viewed as a distributed
quantity induced by the uncertainty of input parameters \cite{le2010spectral}, rather than as a deterministic quantity.  
When the input parameters themselves are spatially (and/or temporally) extended, one is faced with 
much higher-dimensional problems, and indeed distributions over function spaces in principle
\cite{bogachev1998gaussian, matern1960spatial, whittle1963prediction}.   
In the context of Bayesian inference, this leads to the notion of a Bayesian analogue of the 
classical inverse problem \cite{Stuart10,tarantola2005inverse, KS00, biegler2011large}.  
Such problems are enormously
challenging both algorithmically and computationally, and largely motivate the present work.  At the same
time, a very similar problem of big data is recently attracting a lot of attention.  In the former case, even in the hypothetical case 
of full-field measurements, when the amount of data is infinite, the {\it effective} dimension of the data, or the space 
where posterior measure concentrates with respect to the prior, is often quite small with respect to that of the 
underlying parameter of interest, due to smoothing of the forward problem 
\cite{Stuart10, cui2014dimension, law2014proposals, schillings2014scaling}.  
The big data problem directly confronts the case of genuinely high-dimensional posterior distributions, 
i.e., the posterior differs significantly from 
the prior in the whole space 
\cite{scott2013bayes,korattikara2013austerity, minsker2014robust,ghosal2000convergence,maclaurin2014firefly}.

\subsection{Algorithmic introduction}
\label{ssec:algo}

Probability distributions over low-dimensional spaces are straightforward
to represent via the associated probability density.
It is impossible, however, to represent densities in higher than a few dimensions.
But one can 
do something that is 
usually sufficient in scientific utility: one can sample the 
probability distribution 
with Monte Carlo. 
Probability distributions arising from a Bayesian framework introduce another layer of complexity 
in Monte Carlo, 
as typically one can only {\it evaluate} the posterior distribution, up to a normalizing constant, while 
direct sampling methods typically do not exist.  One must resort to methods such as importance or 
rejection sampling or \gls{mcmc} \cite{gilks486markov, gilks2005markov}.  

A primary workhorse of Bayesian computation is \gls{mcmc}.
 A popular and versatile MCMC algorithm is the \gls{mh}, introduced in \cite{MRRT53} and later revised to its current
form in \cite{Has70}.  
The \gls{am} \cite{haario2001adaptive}, and derivatives thereof 
(DRAM \cite{haario2006dram}, ASWAM \cite{andrieu2008tutorial}, 
SCAM \cite{haario2005componentwise}, RAM \cite{vihola2012robust}, etc.), construct proposals based 
on the empirical covariance arising from the current trajectory, i.e., the past samples.
These proposals are perhaps the most versatile, effective, and useful among 
the \gls{mh}-type
algorithms for 
low-dimensional and reasonably well-behaved targets, for example unimodal
up to a dimension of 100.  
As the proposal depends on the chain history, it is no longer Markov, although there
is theoretical work guaranteeing convergence under fairly general conditions 
\cite{andrieu2006ergodicity, roberts2007coupling, saksman2010ergodicity, fort2011convergence,fort2014central}. 
For targets in which the Hessian of logarithm has a strong local dependence, 
gradient-based proposals 
such as the \gls{mala} \cite{rossky1978brownian, roberts1996exponential} 
or the \gls{hmc} algorithm \cite{duane1987hybrid, neal1995bayesian} or their manifold extensions \cite{girolami2011riemann} 
can improve the convergence time, at the cost of providing the gradients, which may be nontrivial to obtain or 
may not even exist. 
It can be shown that such proposals, as well as the random walk (RW) proposal upon which the AM algorithms are 
based, can be derived from the explicit discretization of a certain \gls{sde}.
Based on such diffusion limits, it has been shown that for underlying dimension $d$, 
the variance, or squared step-size, taken by \gls{rw}, \gls{mala}, and \gls{hmc} algorithms must scale as 
$\cO(1/d)$ \cite{GGR97, BRS09, mattingly2011spde}, $\cO(d^{-1/3})$ \cite{RR98, pillai2011optimal},
and $\cO(d^{-1/4})$ \cite{beskos2013optimal}, respectively.  
This naturally translates to decorrelation time of the inverse order, i.e., 
the number of steps required to obtain an almost independent sample is $\cO(d)$, $\cO(d^{1/3})$, and $\cO(d^{1/4})$
\cite{RR01}.
For high dimensional targets, this is naturally impractical, and this has been a limiting factor for the application of these 
algorithms to targets over higher dimensional spaces, although the gradient-based methods 
can still be effective in high dimensions if Hessian information is incorporated efficiently 
\cite{martin2012stochastic, bui2014solving}.  
If a target arising from a Bayesian inverse problem is well-defined in the function-space limit, as it should be, 
then proposals can be designed to respect that limit \cite{Tie}.  
When the problem is discretized, such proposals exhibit a decorrelation time that is independent 
of the refinement of the mesh towards that limit; in other words, independent of the underlying dimension 
\cite{BRSV08, cotter2013mcmc}, or $\cO(1)$.  
Recently the work \cite{law2014proposals} introduced an algorithm that incorporates general operator-weighting,
and in particular Hessian information, into function-space proposals which may be derived from time-inhomogeneous
discretization of the Ornstein-Uhlenbeck SDE.
The work \cite{cui2014dimension} goes one step further, using prior-preconditioned Hessian information 
to adaptively identify the space of posterior concentration, and then 
using empirical covariance information within that low-dimensional space to adaptively precondition  
a time-inhomogeneous discretization of the Langevin SDE.  

In general, the amount of elaborate forward simulation code in the world, 
whether it be high-dimensional \gls{ode} or \gls{pde}, far outweighs the 
associated gradient and adjoint codes, so often such information may not be available.  
Indeed the possibility of avoiding the person-hours required to construct such code is therefore highly valuable, 
and provides good motivation for constructing {\it non-intrusive}, {\it black-box}, or gradient-free algorithms.
This work presents an alternative approach to those described above, 
in an attempt to combine the best of the worlds above without resorting to gradient information. 
Indeed, the \gls{pcn} proposal of \cite{cotter2013mcmc} arises from a Crank-Nicolson
discretization of an Ornstein-Uhlenbeck \gls{sde} which preserves a certain Gaussian measure.  
In contrast, the \gls{rw} proposal arises from an Euler-Maruyama discretization of a diffusion which
spreads mass to infinity and has no invariant measure.  
It is this property that provides the $\cO(1)$ decorrelation 
time of the former versus 
the $\cO(d)$ of the later. 
From this viewpoint, the advantage of the former is clear even in the absence of a function-space limit.  
Herein we construct a proposal inspired by the \gls{pcn} that preserves a distribution proportional to
the empirical Gaussian obtained from past samples, yielding an asymptotically dimension-independent adaptive 
Metropolis algorithm, which will be abbreviated DIAM.  
That is, the decorrelation time is expected to scale as $\cO(1)$ for reasonably well-behaved distributions, 
and this can be proven for the Gaussian case.
Nonetheless, this will result in a gain of only $\cO(d^{1/2})$ in convergence time for \gls{rmse} quantities.
Therefore, the value is still limited as long as one is limited to $d\leq100$.  
On the other hand, when the dimension of the target becomes much larger, 
the cost of adaptation itself may become a limiting factor due to the required linear algebra.  
The computational contribution consists of mitigating this effect.


\subsection{Computational introduction}
\label{ssec:comp}

From the computational 
perspective, the fundamental 
limiting operations that 
comprise the \gls{am} algorithm, and the \gls{diam} extension
proposed here, are Level 2 and 3 \gls{blas} operations, scaling traditionally as $\cO(d^2)$ and $\cO(d^3)$,
in particular, dense matrix-vector, matrix-matrix 
multiplication, and Cholesky-based matrix inversion.  
These 
operations 
prevent its use in high dimensions, even given the algorithmic advances outlined in the previous section.  
However, it is shown here that one may impose a lag-time 
of $\cO(d)$ between Cholesky-based matrix inversion,
and hence block updates of the covariance, without 
increasing the required 
number of samples
to convergence.  
The algorithm is thereby immediately reduced to $\cO(d^2)$ rather than $\cO(d^3)$, 
in the sense that the cost to obtain $N$ samples is $\cO(Nd^2)$ (assuming the cost of evaluating the 
logarithm of the unnormalized density is at most $\cO(d^2)$).
It is also feasible to reduce the cost of the algorithm to $\cO(d^2)$
by using low-rank Cholesky updates \cite{dongarra1979linpack, vihola2012robust}. 
It is proposed here to use state-of-the-art 
GPU acceleration of dense linear algebra operations 
within the fundamental operations of the 
\gls{am} and \gls{diam} algorithms. Compute-bound 
operations, i.e., Level 3 \gls{blas} kernels, usually benefit the most
from these hardware accelerators because they are able to stress the floating-point units
with significant data reuse at the high level of the memory hierarchy, and they attain
a decent percentage of the theoretical peak performance of the underlying hardware. Memory-bound
operations, i.e., Level 2 \gls{blas} kernels, are 
however limited by the bus bandwidth and how fast the requested
data can be fetched to the floating-point units, due to {negligible data reuse. 
Accelerators provide much higher bandwidth  compared to standard x86 architecture and, therefore, memory-bound kernels can still be accelerated on such hardware.}
All these assume that the data resides already on the GPU memory,
which is not always the case for current architecture model. Data has to be offloaded from 
the host (CPU) memory to the device (GPU) memory through a thin pipe called the 
Peripheral Component Interconnect Express (PCIe), 
which has an order magnitude lower bandwidth
than the GPU. 
{It is illustrated that by distributing the Level 2 BLAS
operations across the GPU, the quadratic scaling is reduced
by a factor of almost 4, by a combination of the slow data transfer through PCIe,
mitigated by asynchronous processing, and the speed-up of the resultant Level 2 BLAS operations owing to the increased memory bandwidth on the GPU.}


The clock frequency of a single processor of CMOS logic has nearly 
reached its physical limit due to power dissipation constraints. The multicore era 
has permitted the introduction of multiple low-frequency cores on a single chip. 
This trend has been reinforced moving forward with the international
exascale roadmap~\cite{dongarra2011international}, where streaming multiprocessor architectures
(NVIDIA GPUs, Intel Xeon Phi, etc.) composed of lightweight cores will be the norm for future 
exascale systems. The value of brute force concurrent (embarrassing) parallelization is therefore seeing an 
increase in value.  
While traditional Monte Carlo methods enjoy this property, Markov chain Monte Carlo methods do not,
as they are inherently serial in nature.  Nonetheless, one can \textit{a posteriori} justify the merging of concurrent 
parallel chains within the framework of \cite{gelman1992inference,brooks1998general}, 
using the so-called potential scale reduction factor 
(PSRF) as a diagnostic to measure convergence.  
This is the approach to parallelization of AM taken in the recent works \cite{craiu2009learn, solonen2012efficient}, 
although neither work confronts a high dimensional parameter.
In \cite{craiu2009learn} the objective is to sufficiently explore 
the state-space in order to identify a partition for regional adaptation.
In \cite{solonen2012efficient} this approach is used to mitigate the cost
of very expensive forward solves.   
Herein, the approach is proposed as a general parallelization strategy for the algorithm,  
indeed with almost perfect scaling efficiency in terms of time.  
The convergence time of the empirical covariance is decreased by concatenating samples from the 
concurrent chains through periodic synchronization.  This gain makes up for the 
slight slow-down in the collection of a given batch of samples, resulting in effectively strong scaling with respect to convergence time.  
It is shown that this allows black-box sampling of targets over very high dimensions.  
As the focus of this work is the new DIAM algorithm, the principle is illustrated for that algorithm, 
but the same principle is expected to apply to AM.

It should be noted that many more elaborate approaches to parallelization of Bayesian computation have recently emerged, 
including \cite{suchard2010understanding, wilkinson2006parallel, strid2010efficient, craiu2005multiprocess,  lee2010utility, jacob2011using, calderhead2014general}.  For example, the authors in \cite{suchard2010understanding} and \cite{lee2010utility} 
developed a CUDA kernel to tackle the most time-consuming
phase of their MCMC simulation using SIMD parallelizations to run on the massive number of CUDA cores
available on the GPU card. Our numerical algorithm relies on BLAS operations, for which most vendors provide
highly optimized implementations on their hardware (e.g. cuBLAS for NVIDIA). Moreover, our implementation is
portable across a range of vendor hardware, thanks to the legacy of the BLAS library.

It should also be noted that more advanced Monte Carlo methods exist for Bayesian computation, 
such as population-based \gls{mcmc} \cite{geyer1991markov, hukushima1996exchange}, 
equi-energy samplers \cite{kou2006discussion}, 
and sequential Monte Carlo samplers \cite{del2006sequential}.
Such methods are indeed necessary for sampling from very complex multi-modal distributions,
but it should be noted that Metropolis-Hastings algorithms appear {\it within} these algorithms as 
a fundamental component, similarly to the way the \gls{blas} operations appear in the \gls{mh} algorithms
as a fundamental component.  The proposed \gls{diam} algorithm is therefore expected to have a great impact as
a fundamental black-box \gls{mh} algorithm.  

The rest of this paper 
is organized as follows. 
In Section \ref{sec:biphid} the problem of Bayesian inference
in high dimensions is introduced precisely, detailed 
definitions of the baseline and benchmark algorithms are given, 
and finally the concurrent formulation is presented 
as well as the convergence diagnostic for its \textit{a posteriori} justification. 
In Section \ref{sec:res} the algorithms are illustrated by some numerical experiments.
In Section \ref{sec:gpu} 
advanced GPU acceleration techniques 
are introduced, as well as the logistical framework for extending to multiple chains.
Performance results are highlighted in Section \ref{sec:perf} and we conclude
in Section \ref{sec:conclusion}.

%% file: Chapter_2.tex


\section{Bayesian inference in high dimensions}
\label{sec:biphid}

\subsection{General problem formulation}
\label{ssec:gen}

The problem considered here is the following.  
Given a quantity of interest $\varphi:\bbR^d \rightarrow \bbR$, 
estimate its expectation with respect to a probability measure $\pi$
\begin{equation}
\pi(\varphi):=\bbE_\pi(\varphi) = \int_{\bbR^d} \varphi(x) \pi(x) dx \approx \frac1N \sum_{n=1}^N \varphi(x_i), \quad x_i \sim \pi.
 \label{eq:qoi}
 \end{equation}
The notation ``$x\sim \pi$" 
indicates that the random variable $x$ follows the distribution of $\pi$.  
The convergence of the approximation given above is a consequence of the Law of large numbers 
for independent identically distributed (i.i.d.) random variables $x_i$ \cite{RC99}, 
and an extension thereof under an assumption of sufficient decay of correlation \cite{MT93}. 

Let $\eta:\bbR^d \rightarrow \bbR_+$, where $\bbR_+=\{t\in \bbR; t \geq 0\}$, and 
assume $Z:=\int_{\bbR^d} \eta(x) dx <\infty$.  Then $\pi=\eta/Z$ is a probability density,
in the sense that $\pi:\bbR^d \rightarrow \bbR_+$ {\it and} $\int_{\bbR^d}\pi(x) dx=1$.
Assume that given $x\in \bbR^d$, $\eta(x)$ can be readily evaluated, but that there is no
direct method for sampling from $\pi$.  
Probability measures in the present work will always have densities with respect to Lebesgue 
measure, and the same notation will be used both for the measure $\pi:\sigma(\bbR^d)\rightarrow [0,1]$, 
where $\sigma(\bbR^d)$ refers to the {\it sigma algebra} of measurable sets in $\bbR^d$,
and its density $\pi:\sigma(\bbR^d)\rightarrow \bbR^+$ with $\int_{\bbR^d}\pi(x) dx=1$.  This should not cause confusion.

Such a problem often arises in a Bayesian context, in which case one has some 
observation $y$ such that $y|x \sim L(x,\cdot)$, where $L(x,\cdot)$ is the {\it likelihood} 
which gives the distribution of the data $y$ conditional on $x$,
and one knows how to evaluate the density $L(x,y)$ point-wise.
The density of the {\it posterior} distribution of $x|y$ is given by
\begin{equation}
\pi(x) = \frac1Z L(x;y) \pi_0(x), \quad Z= \int_{\bbR^d}L(x;y) \pi_0(x) dx,
\label{eq:bayes}
\end{equation}
where $\pi_0$ is the {\it prior} distribution of $x$ before any observation is made,
$L(x;y)$ is the density associated to the law of $y|x$, and the ``$;$" notation is used to
emphasize that the observation $y\in \bbR^{d_y}$ is 
fixed to a given observed value, while $x$ is allowed to vary \cite{RC99}.

Particular attention will be paid to the case in which $d$ is large.  For example, in the context 
of Bayesian inverse problems, $d\rightarrow \infty$ in principle and it is appropriate to formulate
the problem as the discretization of a limiting measure on a function-space $X$.  
In this case the target is a measure $\mu:X\rightarrow \bbR_+$, $\mu(X)=1$, and 
\eqref{eq:bayes} takes the form
\begin{equation}
\frac{d\mu}{d\mu_0}(x) = \frac1Z L(x;y), \quad Z = \int_{X} L(x;y) \mu_0(dx),
\label{eq:bayesip}
\end{equation}
where $d\mu/d\mu_0$ denotes the {\it Radon-Nikodym derivative} of $\mu$ with respect to $\mu_0$, i.e., the ratio $\mu(du)/\mu_0(du)$ of infinitesimal volume elements at the point $u$. 
A sufficient requirement for the above to be well-defined is that
$c^{-1}<\mu_0(L(\cdot;y))<c$ for some $c\in(0,\infty)$ \cite{Stuart10}.  
This context will not be considered further,
however this is the problem to have in mind when we refer to the $d\rightarrow \infty$
limit for Bayesian inverse problems.

The case of big data may also come increasingly to fit into this scenario.
While it has come to refer in the statistics community to the case of large $d_y$ 
\cite{korattikara2013austerity, minsker2014robust}, 
which need not imply large $d$, it would be natural to try to explain high-dimensional data in terms of 
a high-dimensional parameter.  This may again lead to a posterior distribution over a 
high-dimensional space.  For example, in the context of regression, access to 
an increasing number of observations and potential covariates may inspire one
to consider an increasing number of covariates as well as an increasing number of 
observations.  In the Bayesian inverse problem 
context, the data may often be given as a noisy observation of the solution of 
a \gls{pde} with the parameter as input, and the intrinsic smoothing property which 
provides well-posedness of \gls{pde} may hence reduce the effective dimension of
the data even in the case of full-field measurements when $d_y \rightarrow \infty$.
In the big-data context, on the other hand, the data may be genuinely informative
over increasingly high-dimensional parameter spaces which can lead to higher
{\it effective dimension} of the posterior with respect to the prior in comparison with 
the Bayesian inverse problem, albeit with a generally much simpler forward model 
connecting the parameter to the observations.  The general black-box methods 
developed here are expected to be effective in both cases and more.

\subsection{Markov chain Monte Carlo}
\label{ssec:mcmc}

Introduce a Markov chain with transition kernel $\cK:\bbR^d \times \sigma(\bbR^d) \rightarrow \bbR^+$. 
Let $\cP(\bbR^d)$ denote the set of probability densities over $\bbR^d$, 
i.e., functions 
$p:\rightarrow \bbR^d \rightarrow \bbR_+$ 
such that 
 $\int_{\bbR^d}p(x) dx=1$.  By the definition of Markov kernel, for $q\in \cP$, one has that 
 $p(y) = \int_{\bbR^d} q(x) \cK(x,y) dx  \in \cP$. 
 The following short-hand notation is therefore commonly used 
 $p = q \cK$, while the equation 
 $p(\varphi) = \int_{\bbR^d} \int_{\bbR^d} q(x) \cK(x,y) dx \varphi(y) dy$ inspires the analogous
 notation $f = \cK \varphi = \int_{\bbR^d} \cK(x,y) \varphi(y) dy$, 
 so that $p(\varphi) = (q\cK)(\varphi) = q(\cK \varphi) = q(f)$.  
 The unfamiliar reader can think of the discrete state-space analogy of row vectors
 representing probability distributions, column vectors representing 
 quantities of interest, and the transition kernel given by a row stochastic matrix.
A density $\pi$ such that $\pi=\cK\pi$ is referred to as (the density of) an invariant measure, 
and a sufficient condition is reversibility
\begin{equation}
\int_{\bbR^d\times\bbR^d} \pi(dx) \cK(x,dx') = \int_{\bbR^d\times\bbR^d}\pi(dx') \cK(x',dx).
\label{eq:reverse}
\end{equation}
Under additional assumptions of irreducibility and aperiodicity, one has ergodicity of the 
chain, i.e., $\lim_{N\rightarrow\infty} | \cK^N(x_0, \cdot) - \pi |_{TV} = 0$ for any $x_0 \in \bbR^d$, 
and rates can be derived depending essentially on the rate of decorrelation of the chain.  
A consequence of this is that if one sets $x_n \sim \cK^n(x_0,\cdot) = \cK(x_{n-1},\cdot)$, 
then $x_n$ is distributed approximately according to the target $\pi$, hence such $\{x_{n-M}\}_{n=M+1}^{N+M}$
can be used in the approximation \eqref{eq:qoi}.

Indeed if $x_M \sim \pi$ and the {\it autocorrelation function} (ACF) 
$\rho_n := \bbE[x_{m+n}-\bbE(x)][x_m-\bbE(x)] / (\bbE[x-\bbE(x)]^2) = \rho^n$
for some $\rho \in (0,1)$, then a simple calculation shows that 
\begin{equation}
\bbE_{\prod_{n=1}^N (\pi \cK^{n-1})} \Big | \frac1N \sum_{n=M+1}^{N+M} \varphi(x_n) - \pi(\varphi) \Big |^2 \leq \frac1N {\bbE_\pi[x-\bbE_\pi(x)]^2 (1+2/(1-\rho))},
\label{eq:mse}
\end{equation}
where the geometric series identity $\Theta = \sum_{n=1}^{\infty} \rho^n = 1/(1-\rho)$ was used to simplify
the {\it integrated autocorrelation time} (IACT) $1+ 2\Theta$.  Notice that by comparison to the celebrated 
Central Limit Theorem \cite{RC99} for i.i.d. draws, the {\it effective} sample size of the correlated ensemble, 
with respect to the i.i.d. case, may be defined as $N_{\rm eff} = N/(1+2\Theta)$.
   
The Metropolis-Hastings (MH) algorithm, introduced in \cite{MRRT53} 
and refined to its present version in 
\cite{Has}, is perhaps the most popular and versatile 
amongst the \gls{mcmc} methods.  It states
that an essentially arbitrarily chosen transition 
kernel $\cQ$ \cite{Tie} can be composed with an accept/reject
step as follows in order to satisfy reversibility \eqref{eq:reverse} 
with respect to $\pi$.  Given $x_n$, the next sample
$x_{n+1} \sim \cK(x_n,\cdot)$, where the kernel $\cK$ is defined as follows
\begin{itemize}
\item Let $x' \sim \cQ(x_n,\cdot)$,
\item Let \begin{equation}
x_{n+1} = \Bigg \{ \begin{array}{cc} 
x' & {\rm w.p.} ~ \min\{1,\alpha(x_n,x')\} \\
x_n & {\rm else},
\end{array}
\end{equation}
\end{itemize}
 where the {\it acceptance probability} $\alpha$ is defined as 
 \begin{equation}
 \alpha(x_n,x') = \frac{\pi(x')\cQ(x',x_n)}{\pi(x_n)\cQ(x_n,x')}.
 \label{eq:accept}
 \end{equation}
There are clearly infinitely many
possible choices of $\cQ$, 
which leads to a wide range of behaviors of the associated kernels $\cK$.
Essentially one aims to minimize the correlation between the subsequent samples,
which in turn results in a smaller $\rho$ in \eqref{eq:mse} above and hence 
smaller $\Theta$ and larger effective sample size $N_{\rm eff}$.
The Metropolis-Hastings algorithm is ubiquitous, not only as a method in its own right, 
but also as a fundamental component for many other Bayesian computation algorithms,
as mentioned at the end of Section \ref{sec:intro}.

\subsection{Advanced Metropolis-Hastings proposals}
\label{sec:advmet}

This subsection will focus on the \gls{mh} algorithm introduced in the previous 
subsection.  The most basic Metropolis-Hastings proposal will be 
introduced (indeed, the Metropolis algorithm), followed by the more advanced black-box, 
or gradient-free, algorithms which were mentioned in Sec. \ref{sec:intro}.
Finally, the algorithm introduced in the present work will be defined.

\subsubsection{Random Walk}
\label{sec:rw}

The presentation begins with the \gls{sde}
\begin{equation}
dx = A dW
\label{eq:rwc}
\end{equation}
where $A\in \bbR^{d\times d}$ is positive definite 
and $dW$ is an independent increment of Brownian
motion $dW \sim N(0, dt\times I)$ \cite{Oksendal98}.
An Euler-Maruyama discretization
of this equation with step-size $\beta$ (time-step $\beta^2$)
gives \cite{Kloeden92}
\begin{equation}
x_{n+1} = x_n + \beta A W_n,
\label{eq:rw}
\end{equation}
where $W_n\sim N(0,I)$ and $W_n\perp W_m$ for all $n,m$.
The standard random walk (RW) is defined by the above equation
so that 
$\cQ(x_n,x_{n+1}) = \cQ(x_{n+1},x_{n})\propto \exp \{-\frac1{2\beta^2} |A^{-1} (x_n-x_{n+1})|^2\}$.
The fact that the proposal density is symmetric means that $\alpha(x,x')=\pi(x')/\pi(x)$.
Often $A=I$ is chosen as the identity matrix, although it is possible to make other 
educated choices, for example the prior covariance in a Bayesian context, 
the Hessian close to the maximizer,
 or some other 
approximation of the covariance of the target.

\subsubsection{Preconditioned Crank-Nicolson}
\label{sssec:pcn}

In turn, the Ornstein-Uhlenbeck process is defined by the following \gls{sde}
\begin{equation}
dx = -  B x dt + \sqrt{2B} A dW,
\label{eq:ouc}
\end{equation}
where $A$ is as above, $B$ is symmetric and positive definite, $\sqrt{B}$ 
denotes the symmetric matrix square root, and it is assumed that $BA=AB$.
It can be shown that the above equation has invariant distribution $N(0,AA^\top)$,
making it a reasonable equation to aim to approximate if $AA^\top$ is a good
approximation of the covariance of the target.
It was proposed in \cite{BRSV08, cotter2013mcmc} to use the above \gls{sde} as a starting point
with $A=\sqrt{C}$ and $B=I$ for posterior measures with Gaussian prior $N(0,C)$, 
and furthermore to use a Crank-Nicolson discretization scheme, 
leading to the following update, for time-step $\delta$ (upon multiplication by 2):
 \begin{equation}
\big (2+\delta\big)x_{n+1} = \big(2-\delta\big)x_n + 2\sqrt{2\delta} A W_n.
\label{eq:cn}
\end{equation}
Setting step-size $\beta=2\sqrt{2\delta}/(2+\delta)$ one has the 
pre-conditioned Crank-Nicolson (pCN) proposal \cite{cotter2013mcmc}
\begin{equation}
x_{n+1} = \sqrt{1-\beta^2} x_n + \beta A W_n,
\label{eq:pcn}
\end{equation}
with $W_n$ as above.
Notice that this equation {\it preserves} the measure $N(0,AA^\top)$, just like 
its continuum counterpart \eqref{eq:ouc}.  
This means if $p$ is the density of $N(0,AA^\top)$
then $p=p\cQ$, which in turn implies $p(x)\cQ(x,x')=p(x')\cQ(x',x)$.  
So, if $\pi(x)=q(x)p(x)$ for some $q$, then the MH algorithm with this proposal 
has the following acceptance probability $\alpha(x,x') = q(x')/q(x)$.  This is useful 
in case the prior is Gaussian, as only the likelihood appears in the acceptance.
There is nothing intrinsically finite-dimensional about \eqref{eq:ouc}, or its
temporal discretization \eqref{eq:pcn}, so one can see how this allows the definition
of a {\it function-space} algorithm, i.e., one which is defined in the limit $d\rightarrow \infty$
for targets of the form \eqref{eq:bayesip} in which $\mu_0$ is Gaussian.  
Indeed as long as one can construct a proposal which is reversible with respect to the
prior, then the same theory extends to non-Gaussian prior \cite{vollmer2013dimension}.
By observing that the form of \eqref{eq:pcn} may be extended with {\it operators} $B$
replacing the scalar $\beta$, the work of \cite{law2014proposals} introduced general operator-weighted
proposals which are reversible with respect to priors of the form $N(m,AA^\top)$:
\begin{equation}
x_{n+1} = m + A(I-B B^\top)^{1/2} A^{-1} (x_n-m) + A B W_n.
\label{eq:pcnop}
\end{equation}
For the above proposals, Hessian information may be incorporated if it is available,
and this was the strategy of \cite{law2014proposals}.
This was extended to more general proposals including also gradient information, 
and given the general name of dimension-independent likelihood-informed (DILI) 
proposals in \cite{cui2014dimension}.  The name derives from judicious incorporation 
of the linear subspace where the posterior concentrates with respect to the prior, 
the {\it likelihood-informed space} (LIS) \cite{cui2014likelihood}. 

It has been shown in \cite{GGR97} that for proposals of the form \eqref{eq:rw} 
one must have $\beta^2 = \cO(1/d)$, thereby leading to a decorrelation-time of $\cO(d)$.
In turn, by virtue of being defined in the function-space limit, the proposals described
above allow $\beta = \cO(1)$ with respect to parameter dimension.  Of course, the 
effective data dimension, i.e., the dimension of the LIS, 
will indeed still play 
a role for the above proposals, although it can be mitigated for 
DILI proposals, in particular those of the 
type \eqref{eq:pcnop}, by scaling the data-informed directions appropriately.

\subsubsection{Adaptive Metropolis}
\label{sssec:am}

When gradients are unavailable, as assumed in the present work, 
one way to improve upon the proposals 
\eqref{eq:rw} and \eqref{eq:pcn} above is with
empirical covariance information, and this leads to the adaptive Metropolis (AM) algorithm 
\cite{haario2001adaptive}.  Let 
\begin{eqnarray}
\label{eq:sampcov}
C_n & = & \frac{1}{n} \sum_{i=1}^n x_i x_i^\top  -   m_n m_n^\top\\
m_n & =& \frac{1}{n}\sum_{i=1}^n x_i,  
\label{eq:sampmean}
\end{eqnarray}
and choose $A_n$ such that $A_n A_n^\top = C_n$.
Plugging this into \eqref{eq:rw} yields the classical adaptive Metropolis proposal.
The work \cite{GGR97} identifies an optimal acceptance rate of 0.234, and the 
later work \cite{andrieu2008tutorial} proposes to scale adaptively the step-size within the \gls{am} 
algorithm to target such acceptance ratio.  This will be the version of \gls{am} considered here.

\subsubsection{Dimension-independent adaptive Metropolis}
\label{sssec:diam}

The new algorithm introduced here was already alluded to in \cite{law2014proposals, cui2014dimension}.  
This follows naturally from the above presentation by substituting an $A_n$ such that $A_n A_n^\top = C_n$
into \eqref{eq:pcn}.  In fact, a reference point should possibly also be taken into account, 
in which case the proposal takes the form
\begin{equation}
x_{n+1} = x_{\rm ref} + \sqrt{1-\beta^2} (x_n -  x_{\rm ref}) + \beta A_n W_n.
\label{eq:diam}
\end{equation}
The reference point $x_{\rm ref}$ may be chosen as the maximum \textit{a posteriori} (MAP) estimator, i.e., $x_{\rm MAP} = {\rm argmax}_x \pi(x)$, if this is available.  Or else, it may 
be adapted to the empirical mean. \footnote{In this case, it should be set to zero until some sufficiently large
$n$ to avoid too many abrupt jumps of the pivot.}  It is worth dwelling on several points that make this proposal,
and the resultant \gls{mh} algorithm, attractive: 
\begin{itemize}
\item This proposal asymptotically targets $N(u_{\rm ref}, C_\infty)$, which is the best 
Gaussian approximation of the target in case $u_{\rm ref}=m_n \rightarrow m_\infty$, 
for example as measured by Kullback-Liebler (KL) distance \cite{kullback1951information}.  
\item As $\beta \rightarrow 1$, the algorithm converges to the independence sampler.  
Hence, for a Gaussian target, it is easy to see that the acceptance probability approaches 1, following from the previous point.  
\item In the non-Gaussian case, the variance of the proposals will asymptotically coincide with the 
variance of the target, for any $\beta$. 
In turn, the variance of the proposals from the \gls{am} algorithm will be $(1+\beta^2)C_n$.  
So in order for its trace, i.e., the expected $\ell^2$ norm of the \gls{am} proposals, 
to be on par with the target, one will necessarily need to choose $\beta^2 =\cO(1/d)$.  
\item Following from the above, for a Gaussian target, the asymptotic decorrelation-time of the new algorithm
is $\cO(1)$, as opposed $\cO(d)$ for the \gls{am} algorithm.
The new algorithm will hence be called dimension-independent adaptive Metropolis (DIAM).
\end{itemize}

For nonlinear/non-Gaussian targets, it will be necessary to modify the above with some 
additive inflation factor $\alpha>1$ as follows
\begin{equation}
x_{n+1} = x_{\rm ref} + \sqrt{1-\beta^2} (x_n -  x_{\rm ref}) + \beta \alpha A_n W_n.
\label{eq:diamnl}
\end{equation}
Notice that as long as $\alpha \in [1,\sqrt{2}]$, 
 the proposal covariance is still smaller
than in the \gls{am} case, and one therefore expects improved performance, although 
some dependence on dimension will then exist.  This point requires further investigation.

\subsection{Concurrent chains}
\label{ssec:conc}

It is relevant to discuss the potential of ``embarrassingly parallel" 
\gls{mcmc}.  This is a controversial topic, since \gls{mcmc} is an intrinsically serial algorithm and 
convergence proofs typically rely on this fact.  Nonetheless, the works \cite{gelman1992inference,brooks1998general}
describe a convergence diagnostic based on running multiple chains and comparing the 
between-chain and within-chain covariances.  
Once this diagnostic indicates convergence, one is then justified \textit{a posteriori} to merge the samples from the different chains.

\subsubsection{Logistics}
\label{sssec:log}

Denote 
$P$ 
chains by $\{x^p\}_{p=1}^P$, and let \textit{k} denote the number of batches which have been done.  
Each of these is run for \gls{symb:M}
intervals of length \gls{symb:nlag} 
and the local first two moments are collected periodically.  
\begin{eqnarray}
\label{eq:mom2p}
S^p_{k,m}& = & \frac{1}{mn_{\rm lag}} \sum_{i=1}^{m n_{\rm lag}} x_i^p (x_i^p)^\top \\ 
\nonumber
& = & \frac{(m-1) n_{\rm lag}}{m  n_{\rm lag}} S^p_{k,m-1} + 
\frac{n_{\rm lag}}{m  n_{\rm lag}}  \sum_{i={(m-1) n_{\rm lag}}+1}^{m n_{\rm lag}} x_i^p (x_i^p)^\top \\
m^p_{k,m} & =& \frac{1}{m n_{\rm lag}}\sum_{i={(m-1) n_{\rm lag}}+1}^{m n_{\rm lag}} x_i^p \\
\nonumber
& = & \frac{(m-1) n_{\rm lag}}{m  n_{\rm lag}} m^p_{k,m-1} + 
\frac{n_{\rm lag}}{m  n_{\rm lag}}  \sum_{i=1}^{m n_{\rm lag}} x_i^p.
\label{eq:mom1p}
\end{eqnarray}
After each $n_{\rm lag}$ update, 
local updates of the global moments are made
\begin{eqnarray}
\label{eq:mom2pg}
S^{p,{\rm glob}}_{k,m}  & = & \frac{k MPn_{\rm lag}}{(kMP + m)n_{\rm lag}} S^{\rm glob}_{k} +
\frac{m n_{\rm lag}}{(kMP+m)n_{\rm lag}} S^p_{k,m}, \\ 
m^{p,{\rm glob}}_{k,m} & = & \frac{k MPn_{\rm lag}}{(kMP + m)n_{\rm lag}} m^{\rm glob}_{k} +
\frac{m n_{\rm lag}}{(kMP+m)n_{\rm lag}} m^p_{k,m}, 
\label{eq:mom1pg}
\end{eqnarray}
followed by a local update of the global covariance
\begin{eqnarray}
\label{eq:cov2pg}
C^{p,{\rm glob}}_{k,m}& = & S^{p,{\rm glob}}_{k,m} - m^{p,{\rm glob}}_{k,m} (m^{p,{\rm glob}}_{k,m})^\top.
\end{eqnarray}
This is used within the individual steps of the algorithm \eqref{eq:diam}.
Then, each time $m=M$, the local samples from the $P$ chains are merged 
into global moments so they can be shared
\begin{eqnarray}
\label{eq:mom2g}
S^{{\rm glob}}_k & = & \frac{(k-1)}{k} S^{\rm glob}_{k-1} +
\frac{1}{kP} \sum_{p=1}^P S^p_{k-1,M}, \\ 
m^{{\rm glob}}_k  & = & \frac{(k-1) }{k} m^{\rm glob}_{k-1} +
\frac{1}{kP} \sum_{p=1}^P m^p_{k-1,M}. 
\label{eq:mom1g}
\end{eqnarray}
At this point, one can compute the global covariance once, or just return the moments 
to the individual chains to continue in parallel.  This procedure can be optimized, but it
is outside the scope of the present work.

\subsubsection{Potential scale reduction factor}
\label{sssec:psrf}

As mentioned above, the potential scale reduction factor (PSRF) convergence diagnostic will be used for \textit{a posteriori} justification of chain merging.  It is defined as follows.
Start 
$P$ chains, with initial conditions which are over-dispersed with respect to the target. 
Define the following within-chain quantities for each $p$ as follows
\begin{eqnarray}
\nonumber
S^p_{MK} & = & \frac{1}{K} \sum_{k=1}^{K} S^p_{k,M},  \\
\nonumber
m^p_{MK} & = &  \frac{1}{K} \sum_{k=1}^{K} m^p_{k,M},  \\
\label{eq:winco}
\end{eqnarray}
Now define the global quantities for $i=1,\dots, d$:
\begin{eqnarray}
B_i & = & \frac{MKn_{\rm lag}}{P-1} \sum_{p=1}^{P} (m^p_{MK} - m^{{\rm glob}}_K)_i^2,  \\
W_i & = & \frac{MKn_{\rm lag}}{(MKn_{\rm lag}-1)P}\sum_{p=1}^{P} (C^p_{MK} )_{ii},
\label{eq:btwco}
\end{eqnarray}
where $C^p_{MK} = S^p_{MK} - (m^p_{MK})(m^p_{MK})^\top$.
The first quantity is referred to as the {\it between-chain} variance, representing
(a factor $MKn_{\rm lag}$ times)
the variance between the means computed in the individual chains.  The second is the average 
within-chain variance across the chains, and is referred to as the {\it within-chain} variance.
These quantities both approximate the variance.  Now define
\begin{equation}
R_i = \frac{MKn_{\rm lag}-1}{MKn_{\rm lag}} + \left(\frac{P+1}{PMKn_{\rm lag}}\right) \frac{B_i}{W_i}.
\label{eq:psrf}
\end{equation}
The PSRF in this $i^{th}$ direction is given by $\sqrt{R_i}$. One expects that $\sqrt{R_i}>1$ and 
clearly one has that $\sqrt{R_i} \rightarrow 1$ as $K \rightarrow \infty$.  
The indicator for convergence is $\sqrt{R_i} -1 \leq {\rm TOL}$, where ${\rm TOL}$ is taken to
be some number smaller than 0.2.  See \cite{gelman1992inference,brooks1998general} for further
details.

%% file: Chapter_3.tex


\section{Numerical experiments}
\label{sec:res}

This section consists of a systematic collection of numerical experiments that present the 
algorithms defined in this paper.  

\subsection{Description of the test cases}

To begin with, several random posterior densities are introduced.  
First a standard normal random matrix $A \in \bbR^{d \times r}$ is generated, and used to construct a
random symmetric matrix $B=AA^\top$.  Such matrix has a spectrum with maximum eigenvalue $\cO(d)$ and minimum 
eigenvalue close to zero ($r=d$) or zero ($r<d$).  
To mimic the case of a posterior distribution, with standard normal prior and log-likelihood 
$-\frac12 x^\top B x$, the target is fixed as $N(0,C)$, where the covariance is set to 
the form $C = (B+I)^{-1}$. 
This covariance has smallest eigenvalue $\cO(1/d)$ and largest close to 1, which will emphasize the effect of anisotropy.
Furthermore, evaluating the target in these cases requires a dense matrix vector multiplication which has a complexity 
$\cO(d^2)$ and is thus greater than or equal to 
the cost of a typical black-box PDE forward solver one may encounter in a more realistic example. 
The following ``twisting" function is introduced 
$$\phi(x)=(x_1,x_2 + b_1 x_1^2, x_3, x_4 + b_3 x_3^2, \dots, x_{d/10} + b_{d/10-1} x_{d/10-1}^2, x_{d/10+1}, \dots, x_d),$$
which allows the construction of simple ``boomerang" shaped targets 
with exactly computable moments.
The following four Gaussian cases and two non-Gaussian cases are considered:
\begin{itemize}
\item $\pi_1 = N(0,C=(B+I)^{-1})$, $r=d$ (full-rank, ${\rm cond}(C)=\cO(d)$),
\item $\pi_2 = N(0,C=(B/d+I)^{-1})$, $r=d$, (full-rank, ${\rm cond}(C)=\cO(1)$),
\item $\pi_3 = N(0,C=(B+I)^{-1})$, $r=d/10$, (low-rank, ${\rm cond}(C)=\cO(d)$),
\item $\pi_4 = N(0,C=V {\rm diag}[(\sigma^{-2} n^{-4} + 1)^{-1}]_{n=1,\dots,d} V^\top)$, 
(full-rank, ${\rm cond}(C)=\cO(\sigma^{-2})$)
\item $\pi_5 = \pi_1 \circ V \circ \phi \circ V^\top $, $b_i= b \sigma_i^{-2}/\sqrt{d}$, 
$b=0.3$ (non-Gaussian, {\it mildly} twisted),
\item $\pi_6 = \pi_1 \circ V \circ \phi \circ V^\top $, $b_i=b \sigma_i^{-2}/\sqrt{d}$, 
$b=2$ (non-Gaussian, {\it strongly} twisted),
\end{itemize}
where $V \Sigma V^\top=C$ is the {\it ordered} eigendecomposition of $C$ from $\pi_1$, 
such that the first eigenpair corresponds to the smallest eigenvalue of $C$.
Notice that the Jacobian determinant of $\phi$ is $1$, so a change of variables is trivial.  
Also, one can compute the maximizer of $\pi_j$ for all $j$ and it is 0.  
Furthermore, the mean and variance of $\pi_5$ and $\pi_6$ 
for $i=2,4,\cdots, d/10$ are given by
\[
\bbE[(V^\top x)_i] = - b_{i-1} \sigma_{i-1}^2, \quad 
\bbE[(V^\top x)_i - \bbE[(V^\top x)_i]]^2 = \sigma_i^2 + 2 b_{i-1}^2 \sigma_{i-1}^4,
\]
where $\sigma_i^2$ are the variances of the $i^{th}$ component under $\pi_1 \circ V^\top$, i.e., the $i^{th}$
diagonal element in $\Sigma$.  For the others, the mean is $0$ and the covariance is $C$, of course.
For the last two non-Gaussian distributions 
$\alpha>1$ must be tuned 
in \eqref{eq:diamnl}, to allow sufficient spread in the proposal.  This is purely heuristic.

The above targets are all randomly generated, but chosen to mimic certain problems that arise in practice. 
We fix a modestly high dimension $d$=100. 
The target $\pi_1$ 
has the structure one might encounter in a big data problem, where we reduce the dimension of the data to $d_y=d$.  This target
is highly anisotropic because the covariance has a big a condition number, 
which may or may not be the case for a big data problem, 
but which makes the problem more challenging. 
The target $\pi_2$ is generated by deliberately reducing the condition number from $\cO(d)$ to $\cO(1)$, thus making a clear comparison with $\pi_1$ to show how condition number impacts 
the algorithm efficiency. The target $\pi_3$ 
simulates the context of a Bayesian inverse problem, in which 
the posterior is low-rank with respect to the prior. The target $\pi_4$ has the structure of 
a Bayesian inverse problem with ``smoothing" forward map, for example from a PDE forward solve, 
given by the decaying spectrum of the likelihood.  
The parameter $\sigma^{-2}$ in this case corresponds to $1/$variance on the data.  
Smaller variance implies bigger condition number, which makes this distribution more anisotropic and thus harder to sample from. 
The targets $\pi_5$ and $\pi_6$ are non-Gaussian distributions: $\pi_5$ is a mildly twisted Gaussian and $\pi_6$ is a strongly twisted Gaussian.

\subsection{Autocorrelation assessment}

\begin{figure}
	\centering
	\includegraphics[width=.75\textwidth]{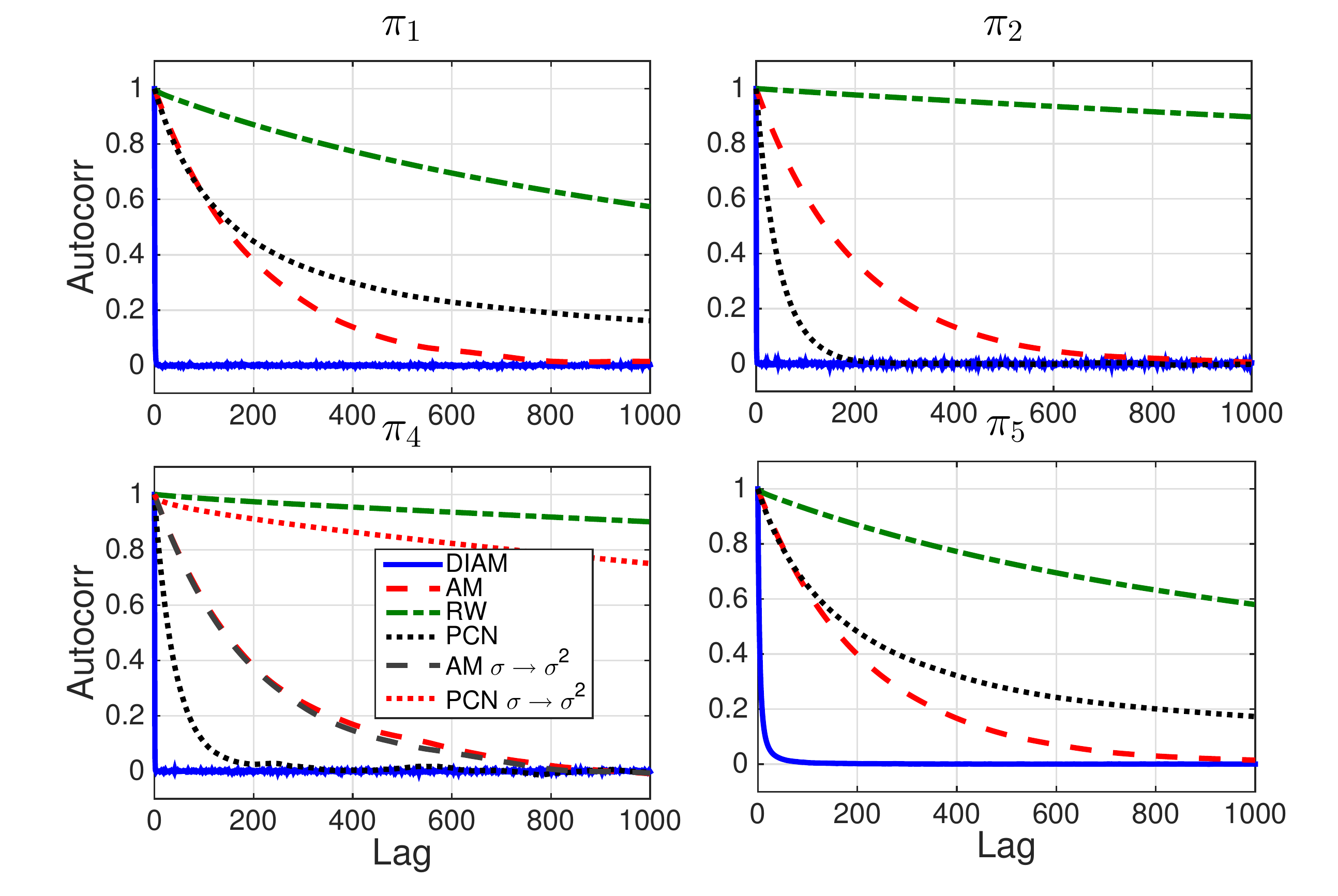}
	\label{fig:acfpi1}
	\includegraphics[width=.75\textwidth]{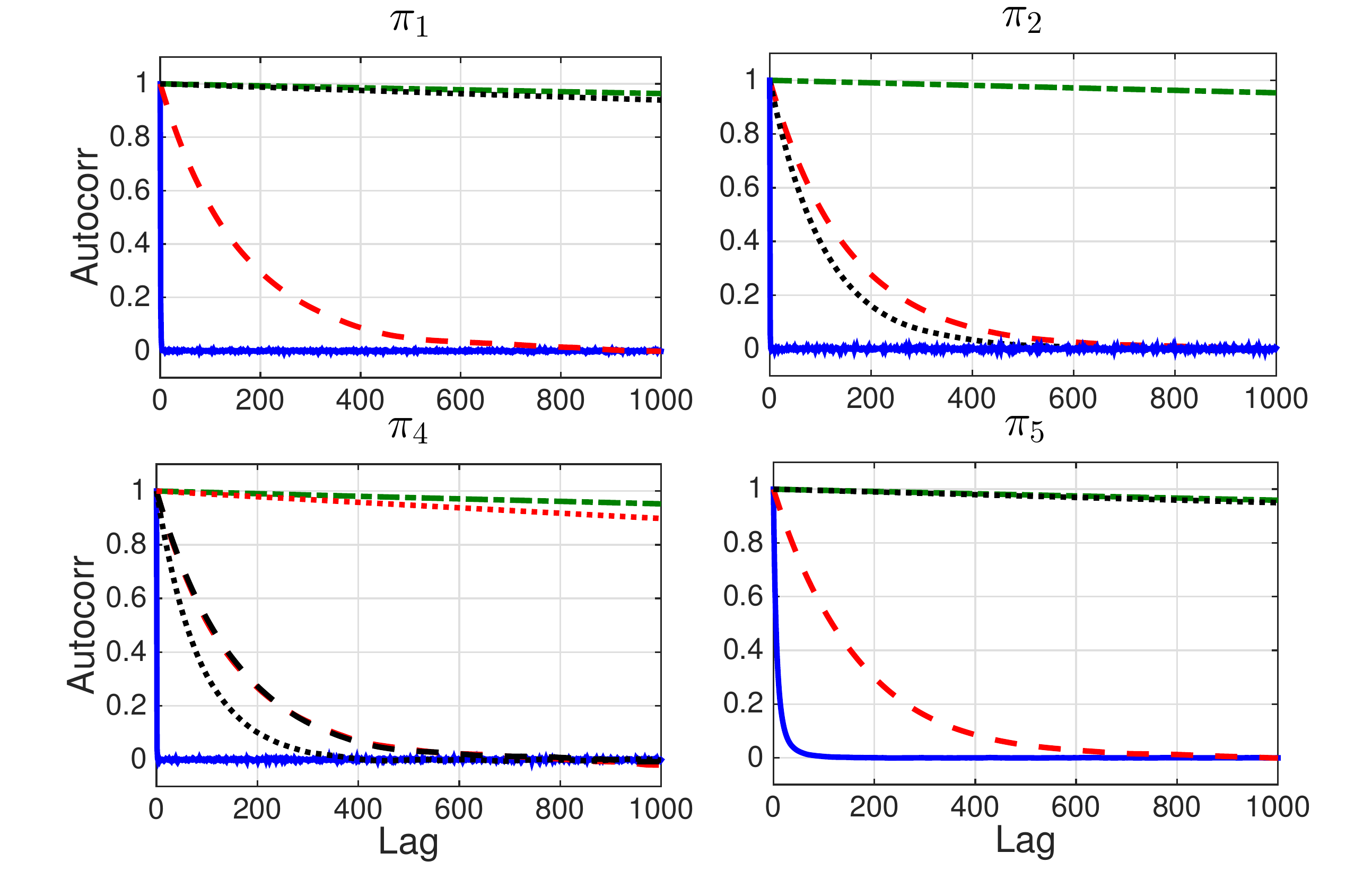}
	\includegraphics[width=.75\textwidth]{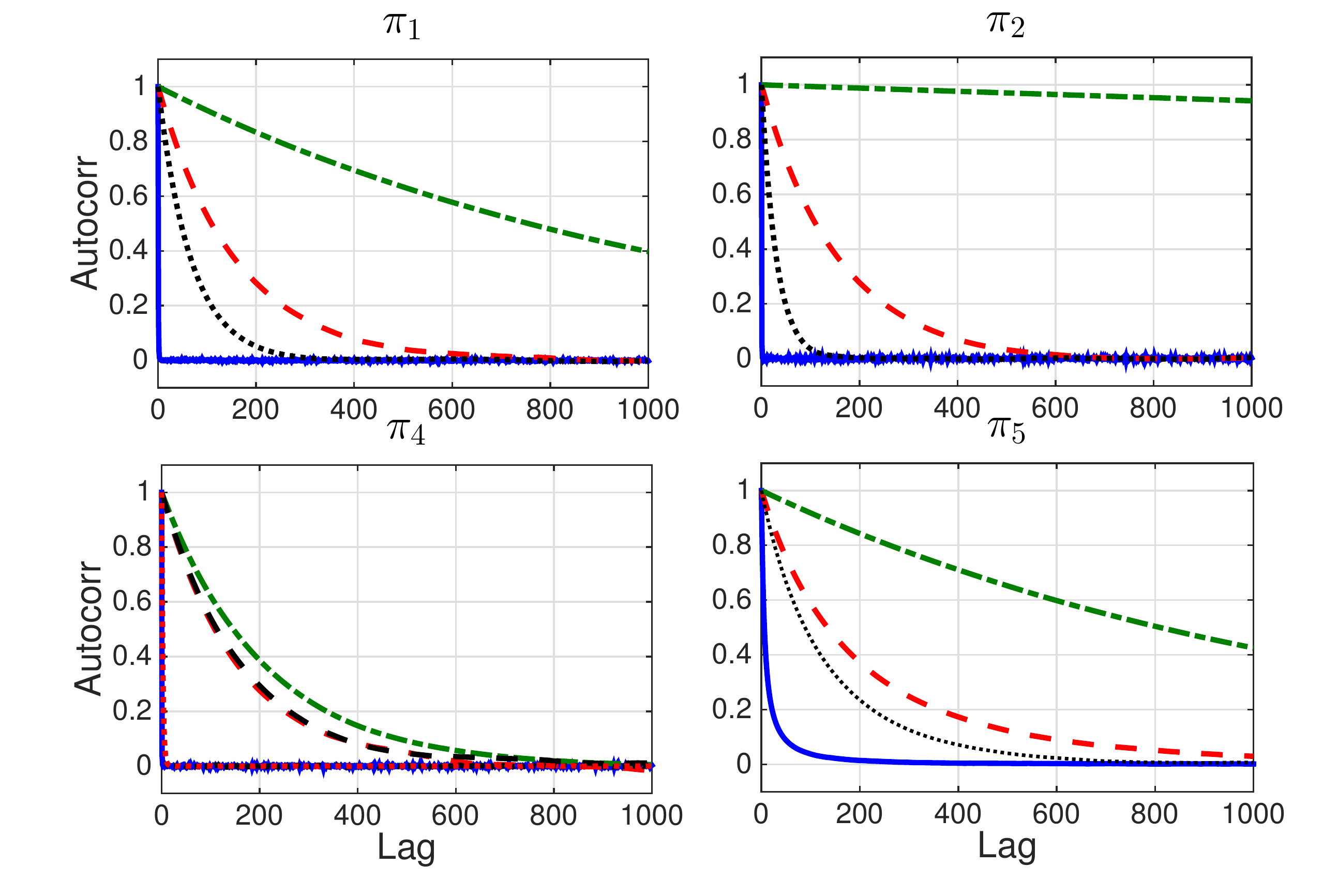}
	\caption{Comparison of autocorrelation function of the log posterior (top four panels), and the projection onto the eigenvector associated with the largest eigenvalue (middle four) and the smallest eigenvalue (bottom four) of DIAM, AM, pCN, and RW on targets $\pi_i$ for $i=1,2,4,5$.}
\end{figure}

In Figure~\ref{fig:acfpi1} 
the numerical performance of the \gls{diam}, \gls{am}, \gls{pcn}, and \gls{rw} 
are compared
by looking at their autocorrelation functions 
with underlying distributions $\pi_1$ through $\pi_6$ for $d=100$. 
The step-size $\beta$ is adapted by targeting the optimal acceptance ratio range, 
which is 0.1 to 0.3 for \gls{am} and \gls{rw} and is 0.3 to 0.5 for \gls{diam} and \gls{pcn}. 
It is chosen initially as $2.4/\sqrt{d}$,  which is suggested 
in 
\cite{haario2001adaptive, gelman1996efficient}.
The top four panels of Figure~\ref{fig:acfpi1} show the autocorrelation function of $\varphi(x) = \log\pi_i(x)$, for $i=1,2,4,$ and $5$, 
as a single global measure of DIAM, AM, pCN and RW. 
The middle and bottom four panels of Figure~\ref{fig:acfpi1} 
show the autocorrelation functions of $\varphi(x)=v_d^T x$ 
and $\varphi(x)=v_1^T x$, the projections onto the eigenvector associated to 
the largest eigenvalue and the smallest eigenvalue, respectively. 
One expects that  \gls{diam} will perform 
the best and \gls{rw} will perform 
the worst. 
The performance of \gls{pcn} and \gls{am} is 
subtle since, on the one hand, \gls{pcn} 
is dimension-independent but isotropic algorithm and may become competitive 
in high-dimensional and well-conditioned cases.  On the other hand, 
the \gls{am} algorithm performs equally in all directions, although suffers from a 
$\cO(d)$ dependence on the dimension, and therefore it 
performs better than 
\gls{pcn} for targets of modest dimension whose covariance has a large condition number. 
The condition number is deliberately increased in target $\pi_4$ from $\cO(d)$ (for $\sigma^{2}=1/d$) to $\cO(d^2)$, 
so that one can have a more clear idea on how AM and pCN react as the condition number of $C$ increases. 
This is shown in the bottom left panels, where there are two curves for each of pCN and AM.
The AM algorithm performs the same.  The pCN algorithm, on the other hand, performs the same for the 
eigen-direction corresponding to the smallest eigenvalue (bottom panels), 
but performs significantly worse on both other functionals.
Numerical experiments confirm the behavior described above.

{
As mentioned, it is expected that the cost of a forward solve, say $C(d)$, 
will be bounded by $\cO(d^2)$, 
so these experiments should give a good measure of the general usability of the algorithm.
For example, if the forward solve 
involves a dense matrix-vector multiplication it is 
$\cO(d^2)$, 
if it involves an iterative solution of a sparse system, it is 
$\cO(d)$, and if it involves
fast Fourier transform (FFT), it is 
$\cO(d\log(d))$.  
The argument found in \cite{law2014proposals}, Section 3.1, indicates that the scaling of pCN is roughly 
$\cO(N\sigma_{\rm min}^{-2}C(d))$, where $\sigma_{\rm min}^{2}$ is the smallest eigenvalue of the posterior covariance, 
with whitened prior (approximately the inverse of the largest eigenvalue of the 
prior pre-conditioned Hessian \cite{flath2011fast, cui2014dimension}), at least in the case of Gaussian targets.  
In turn, AM is 
$\cO(N\max\{d^2,C(d)\}d)$, and DIAM is 
$\cO(N\max\{d^2,C(d)\})$.  
One can therefore conclude that DIAM will outperform AM, and DIAM will outperform pCN if 
$C(d)\gtrsim d^2\sigma_{\rm min}^{2}$.  AM will outperform pCN only if $C(d)\gtrsim d^3\sigma_{\rm min}^{2}$.}

\subsection{Impact of 
$n_{\rm lag}$ choice}

The outcome of any \gls{mcmc} simulation depends, aside the natural variations 
due to random sampling, on the specific way the run is performed. First of all, 
the chain length must be sufficient, and the burn-in has to be dealt with properly. 
In addition, any algorithm contains a number of tuning parameters that may 
decisively affect the results, and the frequency we update our proposal, 
denoted by $n_{\rm lag}$, is one of the parameters that needs to be tuned.

\begin{figure}[h]
\centering
\subfigure[Total required number of samples as a function of $n_{\rm lag}/d$ 
	(normalized by the number for $n_{\rm lag}=d$), 
	to satisfy a given convergence criterion.  
	The missing points for small $n_{\rm lag}$ and larger $d$ 
	correspond to a ``max-time" criterion of 12 hours.]	{\includegraphics[width=0.48\textwidth]{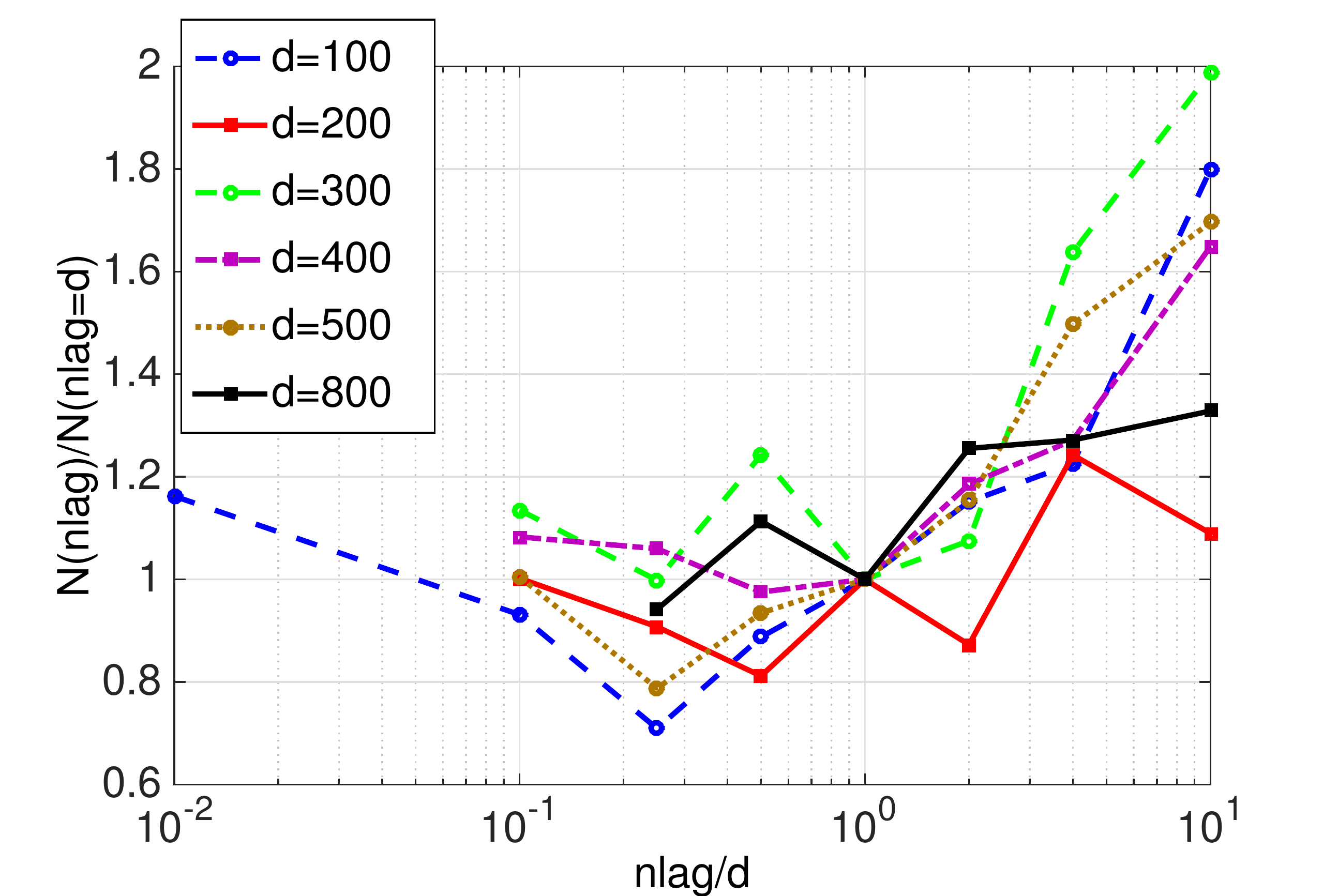}}
	\hfill
\subfigure[PSRF convergence criterion for a range of number of chains $P=4, 10, 16$, 
	for $d=1000$, with the number of outer batch iterations, $k$ given on the x-axis.
	In this case, the chains are stopped when our convergence criterion is satisfied.]
	{\includegraphics[width=0.48\textwidth]{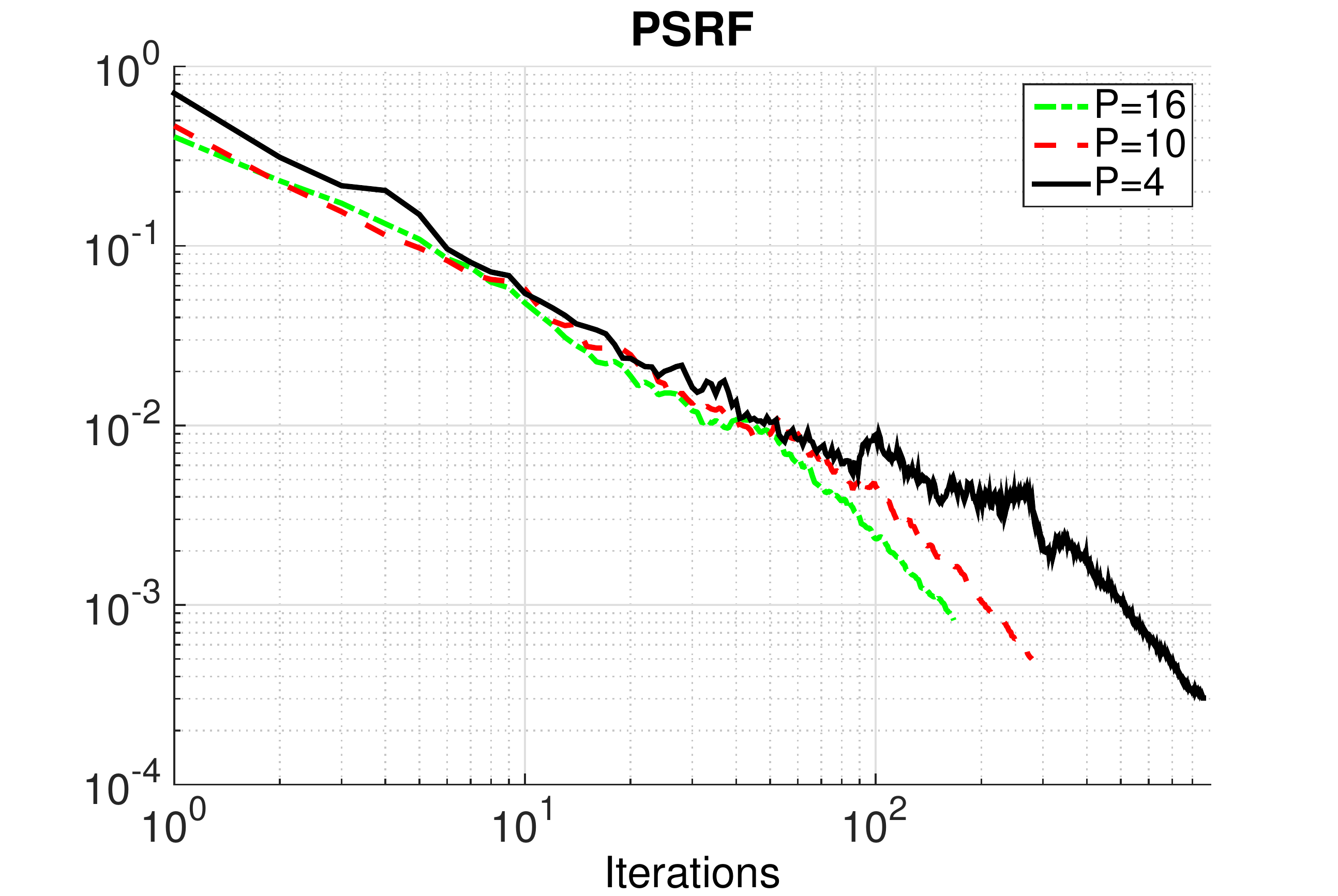}}
\caption{Tuning and convergence diagnostic.} 
\label{fig:snlag_psrf} 
\end{figure}

Test cases with target $\pi_1$ 
were run separately at various values of $d$. For each $d=100, 200, \dots, 500, {\rm and~}800$, 
$n_{\rm lag}$ varied over $\{d/100, d/10, d/4, d/2, d, 2d, 4d, 10d\}$,
and the program was run 
until a certain stopping criterion has been reached. 
The number of samples necessary to reach convergence, normalized by the number 
for $n_{\rm lag}=d$, is shown in the left panel of Fig. \ref{fig:snlag_psrf} as a function of $n_{\rm lag}/d$.
It is interesting that in fact the number of necessary samples {\it increases} for small enough
$n_{\rm lag}$.  
The corresponding time to convergence (not shown) is 
large for either small or large $n_{\rm lag}$, due to the increased number of $\cO(d^3)$ operations in the 
former case and the increased number of required samples in the latter.  
The curves are not convex, although this is presumably due to random effects and 
it is expected that they would smooth out if averages were taken over sufficiently many simulations.
While it would be interesting to identify the optimal value of $n_{\rm lag}$ and see if it converges over multiple
values of $d$, and even targets, to a universal value, for the present purposes this is not necessary.
It suffices to observe that the minimum occurs for some $n_{\rm lag} = \cO(d)$.  
The value of $n_{\rm lag}$ is chosen as $d/2$ in the experiments to follow.  This means that 
the total cost of the algorithm is $\cO(d^2 N)$, where $N$ is the total number of samples.
Similar effect could be obtained by performing low-rank Cholesky updates, although 
Fig. \ref{fig:snlag_psrf} indicates this 
may actually lead to a {\it larger} number of necessary samples to convergence for $n_{\rm lag} = \cO(1)$, 
hence a larger cost. 
{Furthermore, profiling with this choice shows that Level 3 BLAS operations take less than 10\% of the total simulation time,
a consequence of the fact that for sufficiently large $d$, the time to complete $d$ Level 2 BLAS operations of cost $d^2$ is significantly greater than one $d^3$ operation, due to memory constraints.  This is discussed more in the next section.}

For the examples illustrated here, convergence is diagnosed based on the exactly computable moments.
In general, however, such ad-hoc techniques as the potential scale reduction factor (PSRF) described
in Section \ref{sssec:psrf} are required.  The PSRF for $\pi_1$ with $d=1000$ is shown in the right-hand panel 
of Fig. \ref{fig:snlag_psrf} over various $P$, illustrating its convergence.  The convergence criterion that
is used to stop the chains is when the relative error of the sample covariance with respect to the truth
in the Frobenius-norm falls below some TOL.  
{The same convergence criterion, with TOL$= 0.001$,
is used for all the runs except for the tuning of $n_{\rm lag}$.  For the
latter, we use the weaker convergence criterion of the absolute error of 
the sample mean with respect to the truth in the Euclidean norm, with TOL=0.01.}

%% file: Chapter_4.tex


\section{High performance implementation}
\label{sec:gpu}
In this section, we describe the high performance implementation of the DIAM 
algorithm using standard x86 and GPU-accelerated numerical libraries. 

\subsection{Typical CPU-GPU Architecture Ecosystem}
\label{sec:gpucpu}
\begin{figure}[htpb]
	\centering
	\includegraphics[width=\textwidth]{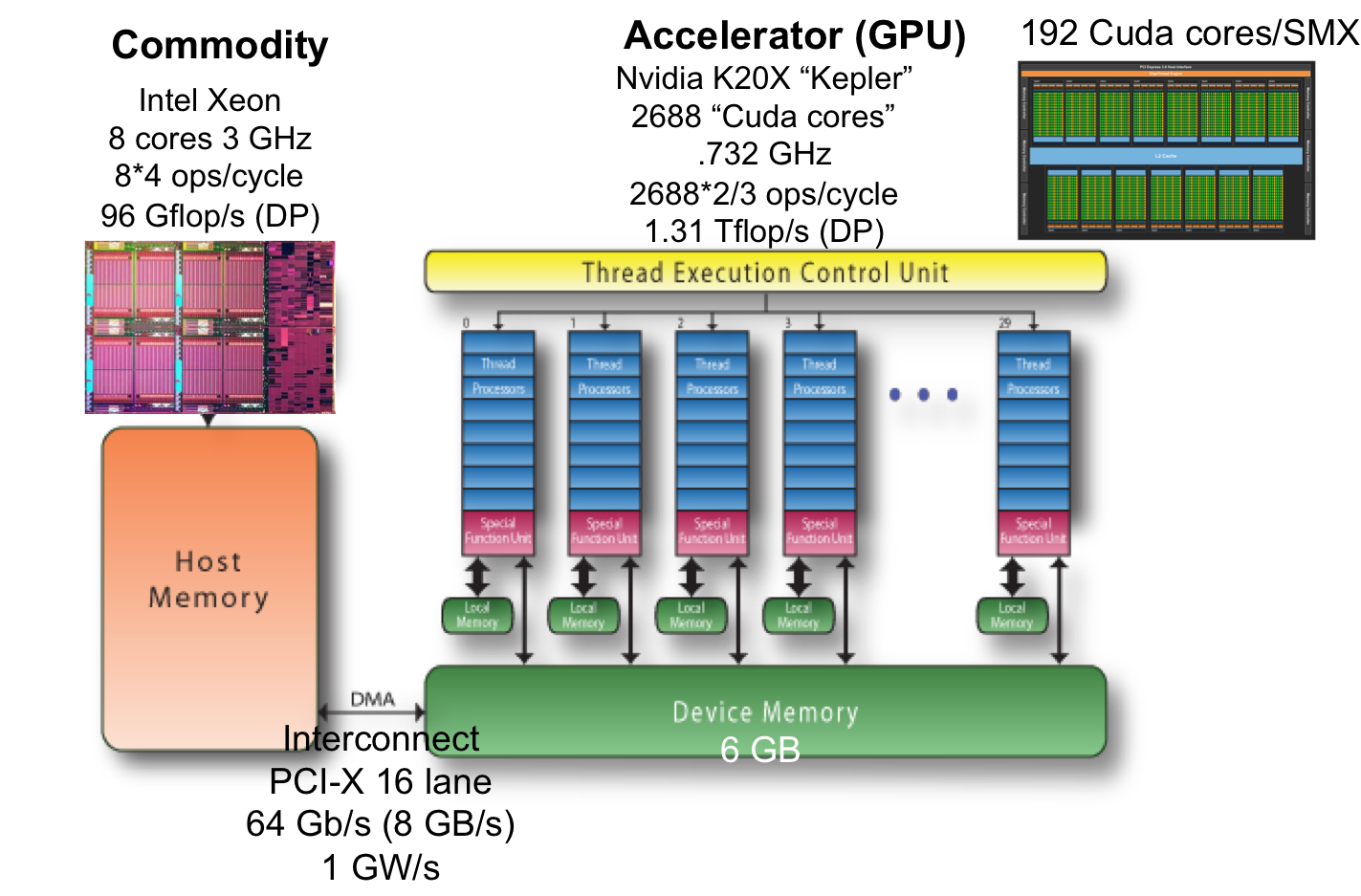}
	\caption{CPU-GPU hardware architecture~\cite{cpugpublockdiagram}.}
	\label{fig:cpugpu}
\end{figure}

Today's hardware landscape is composed of lightweight x86 multicores associated with 
accelerators through a weak link called the Peripheral Component Interconnect Express (PCIe),
as depicted in Figure~\ref{fig:cpugpu}. The architectural discrepancies 
between the host (CPU) and the device (GPU) are manifest. GPU accelerators have 
thousands of CUDA cores, which provide unprecedented parallel performance and computing
capabilities, i.e., more than an order of magnitude higher in terms of 
theoretical peak performance compared to the standard x86 CPU.
Moreover, the speed to fetch data from GPU main memory is higher
{than the standard x86 CPU's bandwidth, by a factor of two or more, depending on the CPU system specifications.  
In our testbed, it is almost a factor of five.} 
However, the PCIe
bus cannot transfer the data from the CPU memory to the GPU memory as fast as the latter can compute.
And this is precisely where the challenge resides, in maintaining the CUDA cores always busy and
not starving for computational work. This problem is further exacerbated 
by the limited size of the GPU memory, which can be 
smaller by one or two orders of magnitude, compared to the CPU memory.
All in all, application performance can usually be leveraged using GPU technology 
(i.e., massive thread parallelism, high computing power and high memory bandwidth)
as long as the overhead of moving data across the PCIe bus can be mitigated by 
using communication-reducing algorithms and/or mandatory communications
{can} be overlapped by useful computations.

\subsection{High performance CPU-GPU numerical software stack}
\label{sec:hpclib}
Fortunately, the high performance numerical software stack targeting the complexity 
of the CPU-GPU hardware is rich in kernel implementations 
and available from optimized open-source and vendor distributions. In particular,
dense linear algebra (DLA) operations are well-supported on multicore and hardware accelerators,
thanks to their regularity in terms of memory accesses. The fundamental DLA kernels are
categorized in three levels: Level 1, 2 and 3, which form the basic linear
algebra subroutines (\gls{blas}) library. Level 1 \gls{blas} involves vector-vector operations (e.g., dot product),
Level 2 \gls{blas} corresponds to matrix-vector operations (e.g., matrix-vector multiplication) and
Level 3 \gls{blas} includes matrix-matrix operations (e.g., matrix-matrix multiplication). While
Level 1 and 2 \gls{blas} operations are mostly memory-bound (limited by the bus bandwidth), Level
3 \gls{blas} kernels are compute-bound thanks to a higher data reuse rate. \gls{blas} kernels are often
at the bottom of the software chain and, therefore, are critical for parallel performance.
Vendors provide support for the \gls{blas} kernels on their respective architectures.  
{F}or instance, Intel
provides its own high performance \gls{blas} library on CPUs, distributed in the Math Kernel Library (MKL)~\cite{mkl}.
On GPUs, NVIDIA provides the cuBLAS library~\cite{cublas}, which implements \gls{blas} kernels using
the CUDA programming model~\cite{cuda}. The open-source KAUST \gls{blas} (KBLAS) 
library~\cite{kblas} provides also a subset of 
Level 2 \gls{blas} operations on GPUs, which performs better than the 
corresponding kernel from NVIDIA cuBLAS. Last but not least, LAPACK~\cite{lapack} provides 
CPU implementations of high-level DLA operations, such as solvers of linear equations and 
covariance (symmetric) matrix inversion.

\subsection{The DIAM software framework}
Below is the work flow of DIAM \footnote{It should be noted that there are other empirical details which are omitted.  
For example, a transient number of initial iterates $n_0$ are collected, with burn-in discarded, 
before the covariance is updated for the first time.} 
for sampling the target 
$\pi : \sigma(\bbR^N) \rightarrow [0,1]$. 
\begin{algorithm}
\caption{DIAM algorithm}
\begin{algorithmic} 
\STATE
\textbf{Initialize} $x_0 
\sim N(0,{\rm Id})$, $A_0=$Id, 
$\beta={\rm min} \{ 2.4/\sqrt{d}, 0.5 \}$, $n=0, n_{\rm accepted}=0$;
\STATE \FOR{convergence criterion $\geq$ TOL} 
\STATE Propose: {$x^* = x_{\rm ref} + \sqrt{1-\beta^2} (x_n -  x_{\rm ref}) + \xi_n;$ }
\STATE $u \sim$ Uniform$(u;0,1)$
\STATE $\log\alpha = \log \pi(x^*) + \frac1{2\alpha^2} (A_n^{-1}x^*)^\top(A_n^{-1}x^*) - 
\left (\log \pi(x_n) + \frac1{2\alpha^2} (A_n^{-1}x_n)^\top(A_n^{-1}x_n) \right)$
\IF{$\log u < \log \alpha$} 
\STATE Accept the proposal: $x_{n+1} 
= x^*$;  $n_{\rm accepted}= n_{\rm accepted}+1$; 
\ELSE
\STATE Reject the proposal: $x_{n+1}
= x_n$ 
\ENDIF
\IF{
$n=k n_{\rm lag}$, $k\in \bbZ$} 
\STATE Calculate acceptance ratio $\hat{\alpha}=n_{\rm accepted}/n_{\rm lag}$ and update $\beta$ 
(increase if \\
$\hat{\alpha}>\alpha_{\rm max}$ or decrease if if $\hat{\alpha}<\alpha_{\rm min}$); 
$n_{\rm accepted}=0$;
\STATE Calculate empirical mean and covariance $m_n, C_n$ as \eqref{eq:sampmean}, \eqref{eq:sampcov}; 
\STATE Update $A_n 
={\rm Cholesky}(C_n)$; Compute $A_n^{-1}$
\STATE {Batch update 
$[\xi_{n+1}, 
\dots,\xi_{n+n_{\rm lag}}] = \beta \alpha A_n [W_{n+1}, 
\dots, W_{n+n_{\rm lag}}]$, \\
where 
$W_m \sim N(0,{\rm Id})$ i.i.d.;}
\ENDIF
\STATE $n=n+1$; 
\ENDFOR
\end{algorithmic}
\end{algorithm}
{There is an evaluation of the log target at each iteration, which is a Level 2 BLAS operation for all of our random targets, 
and another Level 2 BLAS evaluation for the multiplication by $A_n^{-1}$ in the evaluation of the weighted quadratic.  
Every $n_{\rm lag}$ iterations there is a Level 2 BLAS operation for evaluation of the mean, 
and Level 3 BLAS operations for evaluation of the second moment, Cholesky-based matrix inversion, and evaluation of the next 
 $n_{\rm lag}$ random search directions.  
 Nonetheless, the bottleneck with increasing dimension turns out to be the $n_{\rm lag}$ Level 2 BLAS operations in 
 between updates, {given that $n_{\rm lag} = \cO(d)$ and the Level 2 BLAS operations are memory-bound.} 
 Notice $A_{n_{\rm lag}+m}=A_{n_{\rm lag}}$ for $m<n_{\rm lag}$.}
Therefore, from this work flow, DIAM framework is basically composed by the following Level 2 and 3 \gls{blas} 
{operations}: 
\begin{itemize}
\item LARNV: random matrix generation function (auxiliary LAPACK function).
\item TRMV: performs triangular matrix-vector operations (Level 2 \gls{blas}).
\item SYMV: performs symmetric matrix-vector  operation (Level 2 \gls{blas}).
\item GEMV: performs general matrix-vector operations (Level 2 \gls{blas}).
\item SYR: performs the symmetric rank 1 operation (Level 2 \gls{blas}).
\item GEMM: performs general matrix-matrix operations (Level 3 \gls{blas}).
\item POTRF: performs Cholesky factorization (LAPACK function, mostly composed of Level 3 \gls{blas}).
\item POTRI: computes the inverse of a real symmetric positive definite
 matrix $A$ using the Cholesky factorization (POTRF) $A = U^TU$ or $A = LL^T$ 
(LAPACK function, mostly composed of Level 3 \gls{blas}).
\end{itemize}
All these functions are available from the high performance numerical CPU and GPU libraries, 
introduced in Section~\ref{sec:hpclib}.

\subsection{
{Single chain parallelization} implementation challenges}
\label{mpdiamchallenge}
The challenge now resides in composing with all libraries and in determining
which kernels need to run on which platform. Level 2 and 3 \gls{blas} operations 
usually perform best on GPUs, i.e., the Cholesky-based symmetric matrix 
inversion of the sample covariance computation from Equation~\eqref{eq:sampcov}
and the dense matrix-vector multiplication, as highlighted in Equation~\eqref{eq:diam}.
On the one hand, the Cholesky-based matrix inversion is compute-intensive and 
its complexity may impede performance scalability of the overall parallel DIAM 
approach, if frequently requested for solving high-dimension problems.
On the other hand, the dense matrix-vector multiplication is memory-bound and, therefore, 
exhibits a lower arithmetic complexity and slows the 
parallel DIAM 
implementation down if it becomes predominant. The lag-time is then paramount to balance 
these two operations and to further reduce the time to solution, {and it warrants further investigation.}
We rely on existing high-performance implementations of both 
operations: {we use the KBLAS~\cite{kblas} and the NVIDIA cuBLAS~\cite{cublas} libraries
for the Level 2 BLAS operations on GPU occurring each iteration and the Intel MKL library~\cite{mkl} to perform the 
Cholesky-based matrix inversion and other Level 3 BLAS operations occurring once every 
$n_{\rm lag}$ iterations}.
This hybrid CPU-GPU implementation requires the data movement between CPU and GPU memory 
through the slow PCIe link. Ideally, one should try to operate on persistent 
data once on GPU memory to increase data reuse
within the simulation. When this is not feasible, data motion 
has to be hidden using asynchronous
data communication to mitigate the overhead of the slow PCIe bridge. 
The cuBLAS and KBLAS libraries provide API functionalities to ensure communication can be 
overlapped with computation, through the CUDA programming model using the
function CUDA\_MEMCPY\_ASYNC.

\subsection{
{Concurrent chain parallelization} using multithreading}
\label{ssec:conc}

\begin{figure}[hptb]
	\centering
	\includegraphics[width=1.0\textwidth]{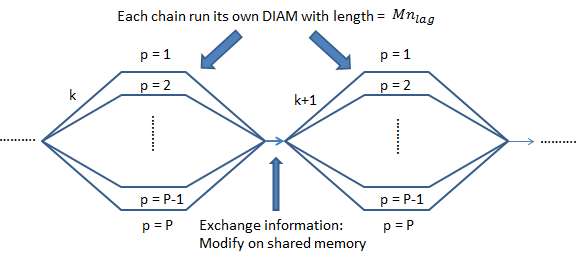}
	\caption{DIAM using the fork and join parallel programming model.}
	\label{fig:forkjoin}
\end{figure}

The degree of parallelism of DIAM can be further leveraged 
by running concurrent chains (see Section~\ref{ssec:conc}).
Thanks to the POSIX threads programming model (Pthreads), threads are instantiated  
and work in an embarrassingly parallel fashion. We rely on the 
usual fork and join parallel programming model to take advantage of the 
parallelism exposed by the concurrent chains. 
Once 
\gls{symb:P} threads are created, each thread $p$ will have its own
private memory containing all needed information to independently process, as depicted in Figure~\ref{fig:forkjoin}. 
In Figure ~\ref{fig:forkjoin}, $k$ denotes the number of batches which have been done.
At the end of each batch processing, 
{the threads are joined using a shared memory lock to facilitate and ensure safe synchronization.
This may engender load imbalance if the workload per thread is not similar. However, 
this can be overcome using a more sophisticated dynamic scheduler to reduce the idle time~\cite{quark}.}

This second level of parallelism introduces another complexity on the CPU because it mixes
threads created by the Intel MKL library (OpenMP) as well as the concurrent chains (Pthreads).
Indeed, MKL implements multithreading in \gls{blas} functions and the default 
number of threads MKL uses corresponds to the number of physical cores 
available on the system, except if the environment variable MKL\_NUM\_THREADS
is defined by the user.
{Thus, the total number of threads running in the system is 
$P\times P_{\rm mkl}$, where $P$ 
is the number of chains launched and $P_{\rm mkl}$ is the number of threads MKL functions fork. 
When 
$P\times P_{\rm mkl}$ is higher than the actually number of cores 
($P_{\rm cores}$) the system has, 
the overall performance may drop down because of thread oversubscription. Therefore, it is 
critical to keep 
$P\times P_{\rm mkl} \leq P_{\rm cores}$.}

%% file: Chapter_5.tex


\section{Performance results}
\label{sec:perf}

This section presents the performance results of various DIAM implementations.

\subsection{Environment settings}
Table~\ref{es} defines the CPU specifications of the computing system used in these experiments. 
Sustained bandwidth is determined by the Stream benchmark. The total number of cores is $20$.

\begin{table}
\vfill
\parbox{.45\linewidth}{
\centering
\label{es}\begin{tabular}{|c|c|}\hline
\textbf{Specifications}&   \\ \hline\hline
 CPU& 2                     \\ \hline
 Cores/CPU & 10                  \\ \hline
 Clock frequency (GHz) & 2.8                 \\ \hline
 Cache size (MB)  &     25                \\ \hline
 Memory Bandwidth (GB/s) & 59.7                  \\ \hline
 Main Memory (GB) & 256                 \\ \hline
 PCI Express & 3.0                 \\ \hline
\end{tabular}
 \caption{Specifications for \textbf{Intel Xeon Ivy Bridge E5-2680 v2}.}
}
\hfill
\parbox{.45\linewidth}{
\centering\label{tscale}
\begin{tabular}{|r|r|c|r|}
\hline
& {\bf Total} &\textbf{Time per} & \\
{\bf $P$} & \textbf{time [s]} & \textbf{batch [s]} & {\bf $PMKn_{\rm lag}$ } \\\hline\hline
 1 & 42517.73                    
  & 5.98 & 142260000\\ \hline
 2 & 22779.92                    
& 8.46  & 107760000\\ \hline
 4 & 9486.76                     
&  8.78 & 86480000\\ \hline
 6 & 5878.11                     
 & 9.09 & 77640000\\ \hline
 8 & 4466.00                     
 & 9.67 & 73920000\\ \hline
 10 & 3506.19                    
 & 9.9 & 70800000\\ \hline
 12 & 3215.66                     
 & 11.01 & 70080000\\ \hline
 14 & 3024.47                     
 & 12.1& 70000000\\ \hline
 16 & 2962.85                     
 & 13.47 & 70400000\\ \hline
\end{tabular}
\caption{Scaling to concurrent chains in terms of convergence time.  Here the total number of samples is 
$N=PMKn_{\rm lag}$.}
}
\end{table}

The system has three NVIDIA Tesla K40 GPU Accelerators with 1.4 TFLOPS sustained performance, 12 GB memory, and ultra-fast 
memory bandwidth 288 GB/s each.  
The machine runs Ubuntu 14.04.1 LTS and provides Intel Compilers Suite 
v13.0 together with the MKL library. The DIAM 
code is written in C
and relies on OpenMP for MKL and Pthreads for the multiple chains implementation
as well as CUDA through cuBLAS and KBLAS, for the CPU and GPU interfaces, respectively.

\subsection{Empirical tuning}

One can notice that Level 3 \gls{blas} functions in DIAM 
are only called every $d/2$ iterations, reducing the algorithm 
complexity to $\cO(d^2)$.
The strategy we use here is that when dealing with small problems, e.g., 
problem sizes smaller than {$1000$}, the optimized Intel MKL~\cite{mkl}, 
is preferred (only CPU), while when dealing with larger problems, e.g., 
problem sizes larger than {$1000$}, high performance libraries such as  
cuBLAS~\cite{cublas} and KBLAS~\cite{kblas} are preferred (GPU).
This tuning choice helps mitigate the overhead of copying data 
between the host (CPU) and the device (GPU).

\subsection{CPU-GPU performance profiling}

{Performance profiling of the MKL-based \gls{diam} CPU implementation indicates that, 
as the dimension increases,} SYMV becomes 
the bottleneck and impedes 
scaling to higher dimensions. 
SYMV is a Level 2 \gls{blas} function and, thus, is limited by the bus 
bandwidth. As described in Section~\ref{sec:gpucpu}, accelerators provide 
{several times} higher bandwidth compared to 
standard x86 architecture and, therefore, memory-bound kernels 
can still be accelerated on such hardware.

\subsection{Performance scalability of DIAM}
\begin{figure}[ht]
\centering
\subfigure[Performance scalability to collect 1e5 samples.]{\includegraphics[width=0.49\textwidth]{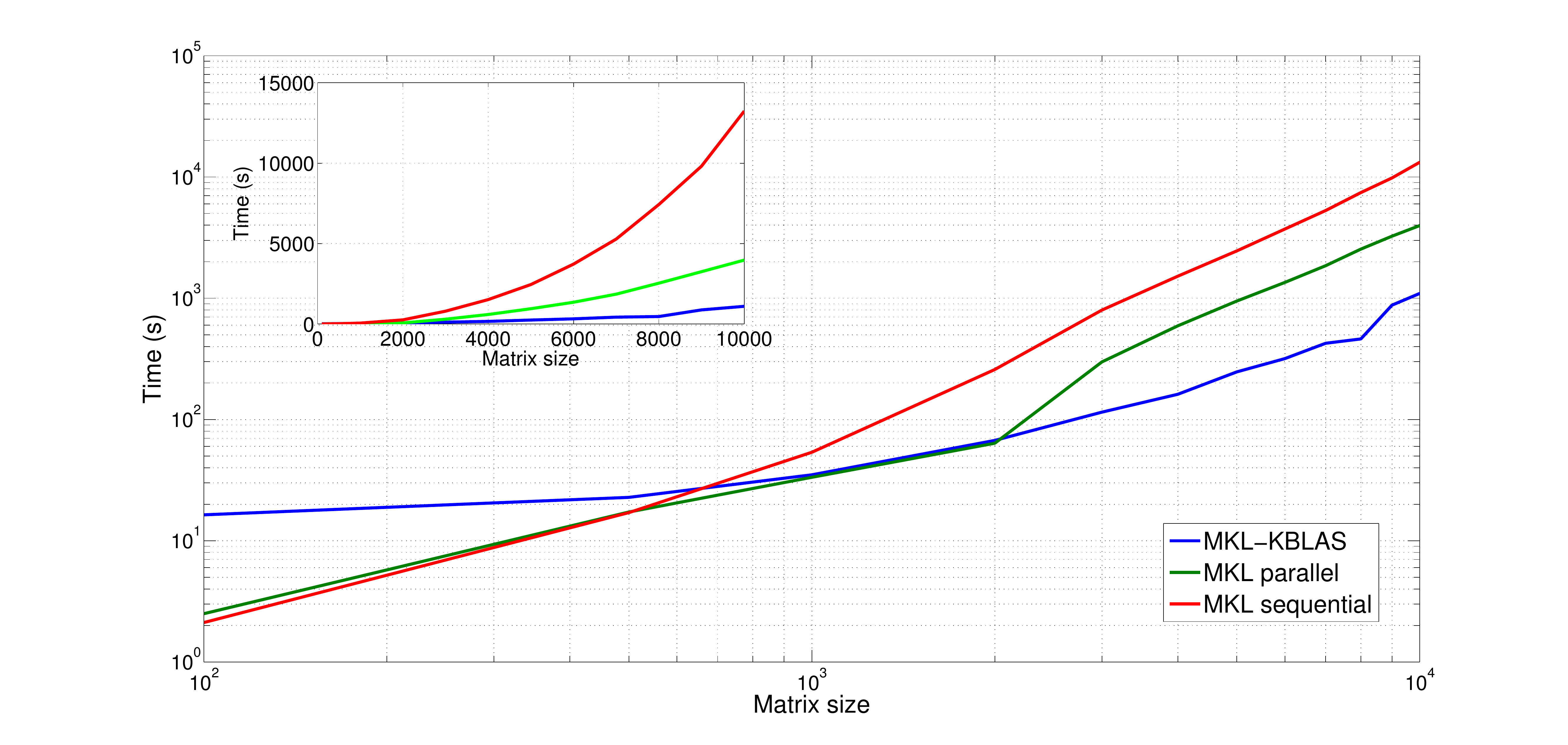}}
\subfigure[Scalability of concurrent chains.]{\includegraphics[width=0.49\textwidth]{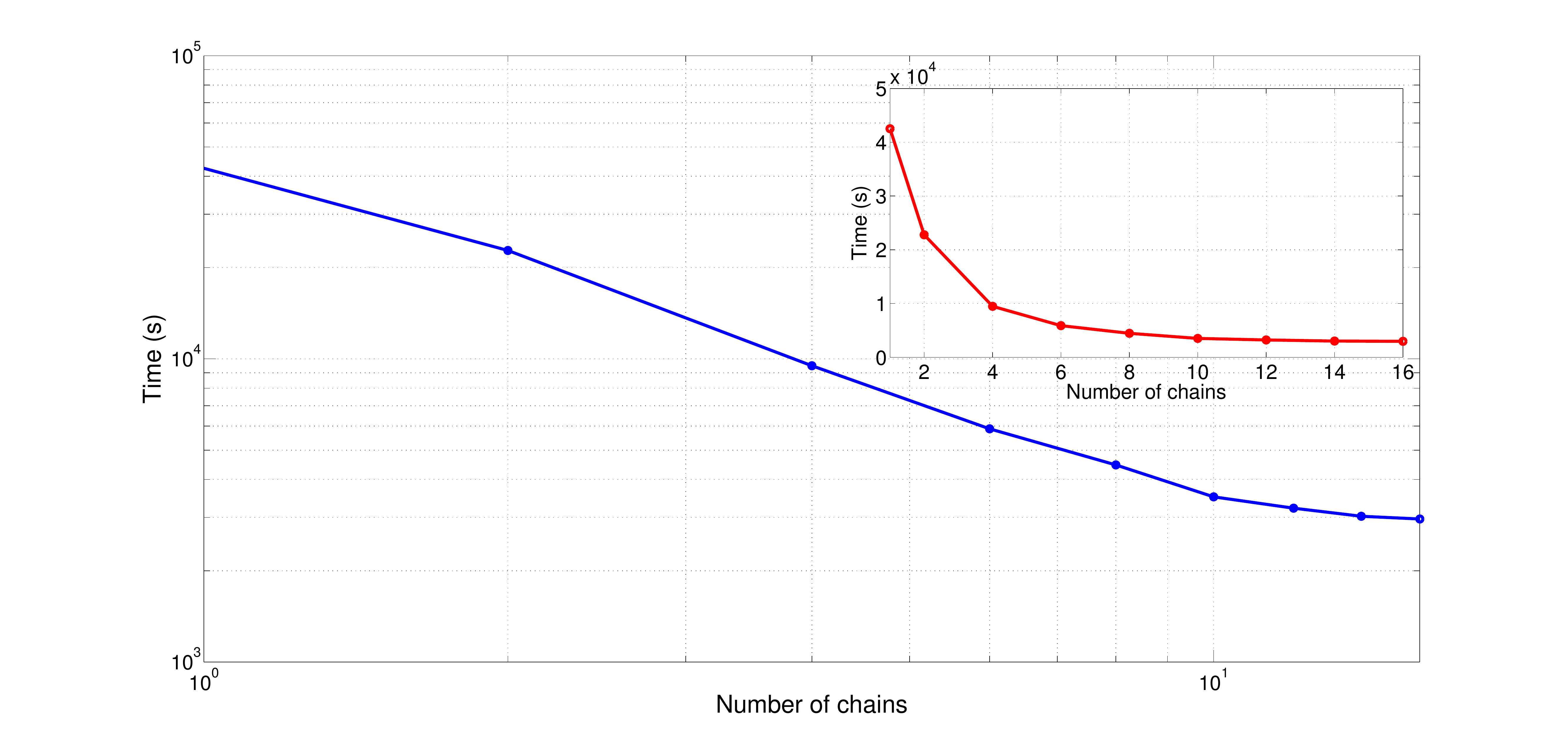}}
\caption{Performance scalability.} 
\label{fig:scaling}
\end{figure}

One of the approaches to statistical inference in high dimensions, 
beside algorithm improvement, is to reorganize the code into a faster implementation. 
In 
Figure~\ref{fig:scaling} (a), we show performance 
scalability in seconds to collect $10^5$ samples from $d=100$ to $d=10000$ using MKL sequential
(by setting MKL\_NUM\_THREADS=$1$), MKL parallel (by setting MKL\_NUM\_THREADS=$20$) 
and MKL-KBLAS (hybrid) high performance libraries combined. 
The target distribution used here is $\pi_1$.
The MKL-KBLAS curve represents 
the implementation using both MKL and GPU-libraries. 
The MKL curve 
represents the implementation only using MKL and run with 20 threads by internally calling OpenMP,  
and MKL sequential 
represents the implementation written on C and run only with one thread with no parallel techniques involved. 
The time required to collect $10^5$
samples of MKL-KBLAS code outperforms that of MKL parallel code 
for {$d\geq 
3000$}.
{Fitting these three curves to quadratic functions results in the following}: 
\begin{itemize}
\item MKL-KBLAS $T=56.39-0.036d+1.34\times10^{-5}d^2$ 
\item MKL Parallel $T=7.49-0.033d+4.32\times10^{-5}d^2$ 
\item MKL Sequential $T=253.63-0.3983d+1.65\times10^{-4}d^2$
\end{itemize}
{These functions make it easy to read the $d_{\rm quad}$ such that quadratic scaling begins, 
as well as the asymptotic gain factor of between $3-4$ in MKL-KBLAS as compared to MKL alone,
and similar between MKL parallel and serial.}

\subsection{Performance scalability of concurrent chain DIAM}

In 
Figure~\ref{fig:scaling}(b), 
{the scaling to concurrent chains is illustrated, {for target $\pi_1$ with $d=1000$, and $M=40$ fixed.}  
The scaling is essentially $T\propto P^{-1}$ at first, but for $P>10$ it slows down, on a machine with 20 cores
(see discussion at the end of \ref{ssec:conc}).
{This algorithm is memory-bound and needs synchronization after each 
chain generates a certain number of samples, thus, 
once the memory 
bandwidth is saturated, adding more threads will have limited benefit 
because 
more time is spent in each batch (the interval between each two synchronization, see Table 5.2). 
We refer to the results on 
Figure~\ref{fig:scaling}(b) ``subtle" 
strong scaling 
because, in contrast to 
the traditional strong scaling, 
the problem size actually is shrinking, namely, the total number of samples required 
to get convergence is decreasing as we add more chains. 
This can therefore still be considered a form of strong scaling because 
the same convergence criterion is used, and in this sense the
problem is the same.  
However, it is clear that the reduction in number of samples is converging. 
The reduction in required number of samples is likely due to the fact that more chains 
translates to more total samples 
used for a given update of 
the proposal covariance, 
hence the proposal adapts faster.
}

{These experiments performed on shared-memory systems suggest new opportunities in further scaling
DIAM to multiple distributed-memory nodes. 
As shown in this section, single-node performance 
starts to decay after running beyond one socket (i.e., ten cores in our testbed) due to the saturation
of the bus bandwidth, which is typical for memory-bound applications. We can then weak-scale the simulation 
by adding more nodes, each equipped with GPUs, and solve higher dimensional problems on a distributed-memory
environment using the Message Passing Interface (MPI)~\cite{mpi}. The synchronization scheme
described in Fig.~\ref{fig:forkjoin} will have to be adjusted and explicit function calls
will have to be made in order to handle communications across the computational nodes. 
In particular, collective communication
operations will be required to synchronize between the distributed nodes. This may generate
overheads due to the higher latency and lower bandwidth of the network interconnect when moving data off-chip.
However, the latest MPI 3.0 standard allows for non-blocking collective communication
operations, which may mitigate the overheads when running on large distributed-memory systems.}

%% file: Conclusion.tex


\section{Summary}
\label{sec:conclusion}

A black-box MCMC algorithm is introduced for Bayesian inference of highly anisotropic targets 
in high dimensions, herein named DIAM.
In particular, it is illustrated that for Gaussian target distributions the integrated autocorrelation time, 
and hence efficiency of the algorithm, is independent of the underlying dimension, asymptotically as the 
number of samples tends to infinity.  The algorithm is illustrated to perform as expected on Gaussian targets,
and also performs favorably with respect to standard AM on non-Gaussian targets.  These algorithms are also 
compared to some other standard Metropolis variants.  {GPU-accelerated Level 2 operations} enable the 
efficient exploration of 
high-dimensional targets with $d \geq 1000$.  The speedup versus standard serial C code
is a factor of twelve as dimension tends to infinity.  
This improvement in conjunction with the combination of 
concurrent chains (justified \textit{a posteriori}) may in principle allow exploration of very high-dimensional targets.
A form of strong scaling with respect to convergence time is illustrated on up to 16 cores.  The parallelization strategy used for 
DIAM 
algorithm will work also for the standard AM algorithm.